\documentclass{amsproc}

\usepackage{
	amsmath, % math style
	amssymb, % twoheadrightarrow required this
	booktabs, % nice tables
	cite, % collect citations, i.e. [3,4,5] -> [3-5] and alike
	multirow, % for text that spans multiple columns/rows
	tikz, % to draw things
}
\usetikzlibrary{decorations.markings, patterns}
\usepackage[pagebackref]{hyperref}
\usepackage[LGR,T1]{fontenc}
\usepackage[utf8]{inputenc}
\usepackage[english]{babel} % language
\usepackage{cleveref} % Overwrites things in hyperref for nice citations. Hence, to be loaded after hyperref
\newtheorem{theorem}{Theorem}[section]
\theoremstyle{definition}
\newtheorem{definition}[theorem]{Definition}
\newtheorem{example}[theorem]{Example}
\numberwithin{equation}{section}

\begin{document}

\title{Root bundles: Applications to F-theory Standard Models}
\author{Martin Bies}
\address{Martin~Bies\\RPTU~Kaiserslautern-Landau\\Department~of~Mathematics\\Gottlieb-Daimler-Straße~48\\67663 Kaiserslautern\\Germany}
\email{bies@mathematik.uni-kl.de}
\subjclass[2020]{Primary }
\date{\today}
\begin{abstract}
The study of vector-like spectra in 4-dimensional F-theory compactifications involves root bundles, which are important for understanding the Quadrillion F-theory Standard Models (F-theory QSMs) and their potential implications in physics. Recent studies focused on a superset of physical root bundles whose cohomologies encode the vector-like spectra for certain matter representations. It was found that more than 99.995\% of the roots in this superset for the family $B_3( \Delta_4^\circ )$ of $\mathcal{O}(10^{11})$ different F-theory QSM geometries had no vector-like exotics, indicating that this scenario is highly likely.

To study the vector-like spectra, the matter curves in the F-theory QSMs were analyzed. It was found that each of them can be deformed to nodal curve that is identical across all spaces in $B_3( \Delta^\circ )$. Therefore, from studying a few nodal curves, one can probe the vector-like spectra of a large fraction of F-theory QSMs. To this end, the cohomologies of all limit roots were determined, with line bundle cohomology on rational nodal curves playing a major role. A computer algorithm was used to enumerate all limit roots and analyze the global sections of all tree-like limit roots. For the remaining circuit-like limit roots, the global sections were manually determined. These results were organized into tables, which represent -- to the best knowledge of the author -- the first arithmetic steps towards Brill-Noether theory of limit roots.
\end{abstract}
\maketitle
\tableofcontents

\newpage
\section{Introduction}

String Theory is a top contender for a theory that unifies quantum field theory and gravity, because it skillfully combines the dynamics of gauge fields with gravity. This means that String Theory must be able to explain all of the various aspects of the physical world that we observe, including the behavior of particles at low energy levels. Most importantly, there is a strong desire to see an explicit representation of the Standard Model of particle physics within String Theory.

A lot of work has been put into achieving this goal using perturbative String Theory, including the $E_8 \times E_8$ heterotic string \cite{CHSW85, GKMR86, BHOP05, BD06, BCD06, AGHL10, AGLP11, AGLP12} and intersecting branes models in type II \cite{BDL96, AFIRU01, AFIRU01*1, IMR01, BKLO01, CSU01, CSU01*1}, as described in \cite{BCLS05} and sources therein. These String Theory compactifications were among the first to produce the Standard Model gauge sector, complete with chiral and, in \cite{BD06, BCD06}, even vector-like spectrum. However, these perturbative models typically suffer from exotic chiral or vector-like matter. The first models which surpassed this hurdle and managed to provide a globally consistent construction of the Minimal Supersymmetric Standard Model (MSSM) using String Theory are \cite{BD06, BCD06}. The works \cite{GLS07, BD08} provide more details on the subtle global conditions for slope-stability of vector bundles.

To accurately describe the gauge dynamics on 7-branes, including their back-reactions on the compactification geometry $B_n$ at all string coupling orders, a non-perturbative extension of type IIB String Theory is required. This can be achieved elegantly through F-theory, as explained in \cite{Vaf96, MV96, MV96*1}. The back-reactions in question are encoded in the geometry of an elliptically fibered Calabi-Yau space, denoted by $\pi \colon Y_{n + 1} \twoheadrightarrow B_n$. The properties of this space can be investigated using established tools of algebraic geometry, which allows for the study and satisfaction of the global consistency conditions of the physics in $10-2n$ non-compact dimensions. For a current overview of the recent developments towards connecting F-theory and the Standard Model, interested readers are directed to \cite{CHST22}.

The chiral fermionic spectrum is a significant feature of particle physics that must be replicated in a realistic 4d $\mathcal{N} = 1$ F-theory compactification, specifically when $n = 3$. In F-theory, the chiral spectrum is determined by a background gauge flux that can be described by the internal $C_3$ profile in the dual M-theory geometry. The chiral spectrum only relies on $G_4 = dC_3 \in H^{(2,2)} (Y_4)$. Several toolkits have been designed to create and count the primary vertical subspace of $G_4$ configurations and identify physical quantities determined by such fluxes \cite{GH12, KMW12, BGK13, CGKP14, CKPOR15, LMTW16, LW16, JTT21, JT22}. These tools have been used to develop chiral F-theory models that are globally consistent \cite{KMW12*1, CKPOR15, LW16, CLLO18, JTT22}. The largest currently-known group of explicit string vacua that realize the Standard Model gauge group with their exact chiral spectrum and gauge coupling unification is the \emph{Quadrillion F-theory Standard Models} (QSMs) \cite{CHLLT19}.

In a 4-dimensional compactification of F-theory with $\mathcal{N} = 1$ supersymmetry, there exist chiral (super)fields in a particular representation $\mathbf{R}$, as well as their charge conjugate counterparts in the representation $\mathbf{\overline{R}}$. The difference between the number of fields in these two representations determines the chiral spectrum. The counting of these fields is done using the vector-like spectrum, which presents additional challenges since the zero modes depend on the potential's flat directions $C_3$ and not just on $G_4 = dC_3$. An F-theory "gauge field" is used to specify the vector-like spectrum and is represented by an element of the Deligne cohomology. The program described in \cite{BMPW14, BMW17, Bie18} relies on the fact that some Deligne cohomology subsets can be expressed as Chow classes. This enables the extraction of line bundles $L_{\mathbf{R}}$ defined on curves $C_\mathbf{R}$ in $B_3$, which can be interpreted as the localization of gauge flux on matter curves $C_\mathbf{R}$ in the dual IIB picture. This in turn renders vector-like pairs on these curves massive. The counting of the zero modes is carried out using the sheaf cohomology groups of the line bundle $L_{\mathbf{R}}$. Specifically, there are $h^{0}(C_{\mathbf{R}}, {L}_{\mathbf{R}})$ massless chiral superfields in the representation $\mathbf{R}$ on $C_{\mathbf{R}}$, while there are $h^1 (C_{\mathbf{R}}, {L}_{\mathbf{R}})$ massless chiral superfields in the representation $\mathbf{\overline{R}}$ on the same curve $C_{\mathbf{R}}$.

The aforementioned process is theoretically sound, but in practice, technical difficulties arise due to the intricate relationship between line bundle cohomologies and the complex structure of $Y_4$. Even with state-of-the-art algorithms \cite{GS23, Dev23, Tea23, **key*} (see also \cite{BMW17, Bie18}) and supercomputers, such as \texttt{Plesken} at \textit{Siegen University}, the necessary computations cannot be performed in realistic compactification geometries. As a result, previous works such as \cite{BMPW14,BMW17,Bie18} focused on models with computationally simple geometries, although they had unrealistically large numbers of chiral fermions. 

Motivated by the progress made in using machine learning techniques for applications in String Theory \cite{CHKN17, HNR19, ACHL21} (see also \cite{Rue20}), recent studies have utilized machine learning techniques in an attempt to overcome this limitation. For instance, the work in \cite{BCDLLR21} investigated the complex structure dependency of line bundle cohomologies in a systematic way. A large dataset \cite{BCDLLR20} of line bundle cohomologies for different complex structure moduli was generated with the algorithms in \cite{aut19}. This dataset was analyzed with with data science techniques. These insights were put on a sound theoretical foundation by use of Brill-Noether theory \cite{BN74}.\footnote{For a contemporary presentation of Brill-Noether theory, we refer the interested reader to \cite{EGH96}. An earlier usage of Brill-Noether theory in the context of F-theory can be found in \cite{Wat16}.} These insights led to a quantitative study of jumps in charged matter vector pairs as a function of the complex structure moduli of the matter curve $C_\mathbf{R}$. 

The quantization condition for a $G_4$-flux -- $G_4 + \frac{1}{2} c_2(Y_4) \in H^4_{\mathbb{Z}}(Y_4)$ -- is often difficult to verify computationally. The authors of \cite{CHLLT19} examined \emph{necessary} conditions for a specific $G_4$-flux to satisfy the quantization conditions. We will proceed under the assumption that the quantization condition is satisfied for the F-theory QSMs. It is important to mention that the proposed $G_4$-flux candidate for the F-theory QSMs cancels the D3-tadpole and renders the $U(1)$-gauge boson massless.

As a result, there is a need to investigate the vector-like spectra present in QSMs, which are localized on five matter curves and were first examined in \cite{BCDLO21}. The study showed that on three of these curves, the line bundle ${L}{\mathbf{R}}$ must be a fractional power $P{\mathbf{R}}$ of the canonical bundle, while on the remaining two curves, the line bundles are altered by contributions from the Yukawa points. These fractional powers of line bundles are known as root bundles, which can be seen as generalizations of spin bundles. The term "P" refers to root bundles throughout the article, inspired by the Greek word  "\begingroup\fontencoding{LGR}\selectfont Ρίζα \endgroup", meaning "root". Similar to spin bundles, root bundles are not unique. The mathematics behind root bundles implies that we should consider different root bundles as being induced from non-equivalent gauge potentials for a given $G_4$-flux.

Typically, we cannot assume that all allowable root bundles on the matter curve $C_{\mathbf{R}}$ can be induced by F-theory gauge potentials in Deligne cohomology. This is similar to how only certain spin bundles on the matter curves are expected to be consistent with the F-theory geometry $Y_4$. Rather, an F-theory gauge potential generates a set ${ P_{\mathbf{R}} }$ that includes one root bundle for each matter curve $C_{\mathbf{R}}$. By repeating this process for all physically relevant F-theory gauge potentials, we should anticipate that only a subset of all root bundles on $C_{\mathbf{R}}$ will generally be reached. Conversely, if we are given a set of root bundles ${ P_{\mathbf{R}} }$ (one for each matter curve), then this set may not originate from an F-theory gauge potential. It is possible for one of the root bundles to not be induced at all or not be induced in combination with one of the other roots.

To determine which root bundles are physical (i.e., induced from an F-theory gauge potential) is crucial for the physics, but it is also a very challenging question. Therefore, \cite{BCDLO21} chose to conduct a systematic study of all root and spin bundles on each matter curve instead of investigating this question. This allowed to name a root bundle and a spin bundle on all matter curves except for the Higgs curve in a particular QSM geometry \cite{CHLLT19}, such that their tensor product is a line bundle with the cohomologies desired for vector-like spectra. Those bundles are found by deforming the smooth, physical matter curve $C_{\mathbf{R}}$ into a nodal curve $C^\bullet_{\mathbf{R}}$ and describing root bundles on the latter in a diagrammatic way using \emph{limit roots} \cite{CCC07}. Counting the global sections of limit root bundles on the full blow-up of $C^\bullet_{\mathbf{R}}$ can be done using the techniques developed in \cite{BCDLO21}.

In \cite{BCL21}, the study was expanded by using a computer algorithm to count all root bundles with three global sections on the full blow-up of $C^\bullet_{\mathbf{R}}$ for various QSM geometries. This required examining families of toric spaces $B_3( \Delta^\circ )$ associated with polytopes $\Delta^\circ$ in the Kreuzer-Skarke list \cite{KS98} (see also \cite{HT17}). These toric spaces are different desingularizations of toric K3-surfaces \cite{Bat94, PS97, CK99, Roh04,BLMSS17}. It was found that the nodal curve $C^\bullet_{\mathbf{R}}$, and thus the number of limit root bundles, only depends on the polytope $\Delta^\circ$. With a computer scan, results were obtained that apply to a large number of QSM geometries. These techniques were refined in \cite{BCDO22}, so that not only full blow-up limit roots but also certain partial blow-up limit roots could be handled. This computer scan indicates that absence of vector-like exotics in the $(\mathbf{3}, \mathbf{2})_{1/6}$ representation is a very likely scenario within the F-theory QSMs. It must be emphasized that this finding has a statistical meaning, as the top-down origin of the limit roots in F-theory gauge potentials has not been clarified yet.

\subsection*{Outline}

This note summarizes and presents the findings of three research works: \cite{BCDLO21, BCL21, BCDO22}. It starts by revising how vector-like spectra in F-theory are related to Deligne cohomology and Chow groups in \cref{sec:Deligne}. Then, \cref{sec:RootBundlesInFTheory} describes the study of root bundles in the context of F-theory, followed by an explanation of nodal deformations and limit roots in \cref{sec:LimitRoots}. In \cref{sec:FRST-Invariance}, it is shown that for F-theory QSMs, the nodal curve is invariant across different geometries, allowing for the approximation of vector-like spectra for a class of F-theory QSMs from a single study. These techniques are then applied to 33 families of F-theory QSMs. A computer scan is used to gauge the statistical likelihood of absence of vector-like exotics, and the cases left by the computer scan are treated manually, as exemplified for the QSM family $B_3( \Delta_4^\circ )$ in \cref{subsec:D4}. Finally, a summary and outlook are given in \cref{sec:SummaryAndOutlook}. The studies in \cite{BCDLO21, BCL21, BCDO22} do not provide a clear picture of which root bundles are realized by F-theory gauge potentials, but statistically speaking, for the selected 33 QSM families, it is likely that many roots are realized in this top-down fashion.

\section{Deligne cohomology and vector-like spectra in F-theory} \label{sec:Deligne}

Type IIB string theory has an effective super-gravity description at the massless level. In flat Minkowski space with 10 dimensions, the effective field theory has supersymmetry $\mathcal{N} = (2,0)$. Its bosonic fields include the dilaton $\varphi$ and the Ramond-Ramond 0-form field $C_0$. The value of the dilaton $\varphi_0$ far away from $D_p$-branes determines the string coupling constant $g_s$ as $g_s = e^{\varphi_0}$ \cite{GSW88, GSW88*1, Pol98, Pol01}. It is convenient to combine the dilaton and the Ramond-Ramond field into a single complex-valued field called the axio-dilaton:
\begin{align}
\tau \colon \mathcal{E}_4 \times B_3 \to \mathbb{C} \, , \, x^\mu \mapsto C_0( x^\mu ) + i e^{- \phi( x^\mu )} \, . 
\end{align}
In this equation, $\mathcal{E}_4$ refers to the internal 4-dimensional real space while $B_3$ refers to the internal 3-dimensional complex space. The axio-dilaton remains constant on $\mathcal{E}_4$. However, due to the backreactions of $D_7$-branes, it varies across $B_3$. Additionally, unless the theory is trivial, the axio-dilaton $\tau$ must become infinite at specific loci of $B_3$, thereby indicating strong-coupling effects in $g_s$.

The type IIB supergravity action has an $\mathrm{SL}(2, \mathbb{Z})$-symmetry that can be used to interpret the axio-dilaton $\tau$ as the complex structure modulus of an elliptic curve $\mathbb{C}_{1,\tau(x^\mu)}$ for each $x^\mu \in B_3$. More information on this can be found in \cite{Wei18} and related works. An elliptic fibration $\pi \colon Y_4 \twoheadrightarrow B_3$ with $\pi^{-1}( x^\mu ) = \mathbb{C}_{1, \tau(x^\mu)}$ serves as a geometric tool that records the value of $\tau$ over $B_3$. Since $\tau$ diverges over certain loci of $B_3$ (unless the type IIB supergravity theory is trivial), this elliptic fibration must be singular. This makes the study of singular elliptic fibrations a common topic in F-theory.

To study F-theory, one could try to work with singular fibrations directly. Among others, \cite{AHK14, CS15, CS15*1, CGSV16, AHKS17} follow this line of thought. However, in this article, the assumption is made that $Y_4$ has a smooth, crepant resolution which is a flat elliptic fibration $\widehat{\pi} \colon \widehat{Y_4} \twoheadrightarrow {B}_3$. Such fibrations may or may not have sections. Studies on F-theory without sections include \cite{Wit96, BM14, AGGK14, KMOPR15, GGK14, MPTW14, MPTW15, MT14, CDKPP15, LMTW16, Kim17}, whereas this article assumes that $\widehat{\pi}$ has a section.

The irreducible components of the codimension-1 locations in $B_3$ where the elliptic curve $\mathbb{C}_{1, \tau(x^\mu)}$ becomes singular are known as gauge divisors. The singularity type over each gauge divisor contains information about an affine Dynkin diagram and thereby relates each gauge divisor to a Lie group. This Lie group is referred to as the gauge group of the corresponding gauge divisor. More details can be found in \cite{Wei18} and references therein.

The irreducible components of the codimension-2 locus of $B_3$ over which the elliptic curve $\mathbb{C}_{1, \tau(x^\mu)}$ is singular, are termed matter curves. The singularity type of each matter curve determines a representation of each gauge group associated to a gauge divisor in which the matter curve is contained. Therefore, matter curves are typically denoted as $C_{\mathbf{R}}$ with $\mathbf{R}$ the representation in question. In quantum field theories, representations are the mathematical counterpart for matter particles. Therefore, it is common to say that matter in representation $\mathbf{R}$ is localized on $C_{\mathbf{R}}$.

Yukawa points refer to the irreducible components of the codimension-3 locus of $B_3$ over which the elliptic curve $\mathbb{C}_{1, \tau(x^\mu)}$ singular. These points are located at the intersections of matter curves. In terms of physics, these points represent interactions among the matter fields that are present on the matter curves which intersect at that point. In F-theory Standard models, these interacts produce the Yukawa interactions in the Standard model.

The $G_4$-flux $G_4 \in H^{(2,2)}( \widehat{Y}_4 )$ needs to satisfy the quantization condition \cite{Wit97}
\begin{align}
G_4 + \frac{1}{2} c_2( T_{\widehat{Y}_4} ) \in H^{(2,2)}_{\mathbb{Z}} ( \widehat{Y}_4 ) = H^{(2,2)} ( \widehat{Y}_4 ) \cap H^{4} ( \widehat{Y}_4, \mathbb{Z} ) \, .
\end{align}
If $\widehat{Y}_4$ is a smooth Calabi-Yau 4-fold that has a globally defined Weierstrass model, then $G_4 \in H^{(2,2)}\mathbb{Z}(\widehat{Y}4)$ due to the evenness of the class $c_2(T{\widehat{Y}_4})$ \cite{CS12}. It can be challenging to verify this condition due to computational limitations. The authors of the \emph{Quadrillion F-theory Standard Models} \cite{CHLLT19} evaluated necessary conditions and confirmed that the chosen $G_4$-flux candidate satisfies both the D3-tadpole cancellation and the masslessness of the $U(1)$-gauge boson.

There is a surjection from the Deligne cohomology group
\begin{align}
\widehat{c}: H^4_D(\widehat{Y}_4, \mathbb{Z}(2)) \twoheadrightarrow  H^{(2, 2)}_\mathbb{Z}(\widehat{Y}_4) \, ,
\end{align}
and a $G_4$-flux can be lifted to an F-theory gauge potential $A$ in the Deligne cohomology group $H^4_D(\widehat{Y}_4, \mathbb{Z}(2))$. While this group is difficult to work with in explicit computations, a subset of it can be parametrized by the map
\begin{align}
\widehat{\gamma}: \mathrm{CH}^2(\widehat{Y}_4, \mathbb{Z}) \rightarrow H^4_D(\widehat{Y}_4, \mathbb{Z}(2)) \, .
\end{align}
Therefore, we focus on investigating $\mathcal{A} \in \mathrm{CH}^2(\widehat{Y}_4, \mathbb{Z})$ that satisfies $(\widehat{c} \circ \widehat{\gamma})(\mathcal{A})= G_4$.

We can link each $\mathcal{A} \in \mathrm{CH}^2(\widehat{Y}_4, \mathbb{Z})$ to a divisor on the matter curve $C_{\mathbf{R}}$ through the cylinder map $D_{\textbf{R}}: \mathrm{CH}^2(\widehat{Y}_4, \mathbb{Z}) \rightarrow \mathrm{CH}^1(C_{\textbf{R}}, \mathbb{Z}) \cong \mathrm{Pic}(C_{\textbf{R}})$. For this, we use the matter surface $S_{\mathbf{R}} \subseteq \widehat{Y}4$ -- actually a 2-cycle and not a surface -- which is a linear combination of $\mathbb{P}^1$-fibrations over $C_{\textbf{R}}$, with the coefficients determined by the weight vector $\mathbf{w}$ of the representation $\mathbf{R}$ (see \cite{Wei18} and references therein). In terms of $\iota_{S_{\mathbf{R}}} \colon S_{\mathbf{R}} \hookrightarrow \widehat{Y}_4$ and $\pi_{S_{\mathbf{R}}} \colon S_{\mathbf{R}} \twoheadrightarrow C_{\mathbf{R}}$, the cylinder map is given by
\begin{align}
D_{\mathbf{R}} \left( \mathcal{A} \right) = \pi_{S_{\mathbf{R}}\ast} \left( \iota_{S_{\mathbf{R}}}^\ast \left( \mathcal{A} \right) \right) \in \mathrm{Pic} \left( C_{\mathbf{R}} \right) \, .
\end{align}

The divisor $D_{\mathbf{R}} \left( \mathcal{A} \right)$ gives rise to the line bundle \cite{BMPW14, BMW17, Bie18}
\begin{align}
P_{\mathbf{R}} \left( \mathcal{A} \right) = \mathcal{O}_{C_{\mathbf{R}}} \left( D_{\mathbf{R}} \left( \mathcal{A} \right) \right) \otimes_{\mathcal{O}_{C_{\mathbf{R}}}} \mathcal{O}_{C_{\mathbf{R}}}^{\text{spin}} \, ,
\end{align}
where $\mathcal{O}_{C_{\mathbf{R}}}^{\text{spin}}$ is a spin bundle on $C_{\mathbf{R}}$ compatible with the global structure of the F-theory compactification on $\widehat{Y}_4$. We will elaborate on $\mathcal{O}_{C_{\mathbf{R}}}^{\text{spin}}$ momentarily. The sheaf cohomologies of $P_{\mathbf{R}} \left( \mathcal{A} \right)$ encode the vector-like spectrum on $C_{\mathbf{R}}$:
\begin{align}
h^0 \left( C_{\mathbf{R}}, P_{\mathbf{R}} \left( \mathcal{A} \right) \right) & \; \leftrightarrow \; \text{chiral zero modes in representation $\mathbf{R}$} \, , \\
h^1 \left( C_{\mathbf{R}}, P_{\mathbf{R}} \left( \mathcal{A} \right) \right) & \; \leftrightarrow \; \text{chiral zero modes in charge conjugate rep. $\overline{\mathbf{R}}$} \, .
\end{align}
Note that the Riemann-Roch theorem implies
\begin{equation}
\chi \left( P_{\mathbf{R}}( \mathcal{A}) \right) = h^0 \left( C_{\mathbf{R}}, P_{\mathbf{R}} \left( \mathcal{A} \right) \right) - h^1 \left( C_{\mathbf{R}}, P_{\mathbf{R}} \left( \mathcal{A} \right) \right) = \mathrm{deg} \left( D_{\mathbf{R}}( \mathcal{A} ) \right) = \int_{S_{\mathbf{R}}}{G_4} \, .
\end{equation}
In the case where there are no unusual vector-like pairs present, the value of $h^0 \left( C_{\mathbf{R}}, P_{\mathbf{R}} \left( \mathcal{A} \right) \right)$ is equal to $\chi(P_{\mathbf{R}} \left( \mathcal{A} \right))$. On the other hand, if there is exactly one vector-like pair, then $h^0 \left( C_{\mathbf{R}}, P_{\mathbf{R}} \left( \mathcal{A} \right) \right)$ is equal to $\chi(P_{\mathbf{R}} \left( \mathcal{A} \right)) + 1$. This second scenario is particularly relevant in the context of constructing an F-theory MSSM, where having precisely one vector-like pair on the Higgs curve is desirable.

We will now discuss the spin bundle $\mathcal{O}_{C_{\mathbf{R}}}^{\text{spin}}$. A smooth genus $g$ curve has $2^{2g}$ spin bundles, so it is necessary to specify which one is being used for $\mathcal{O}_{C_{\mathbf{R}}}^{\text{spin}}$. The cancellation of the Freed-Witten anomaly requires $\text{spin}^c$-structures on $D7$-branes in perturbative IIB compactifications, as shown in \cite{FW99}. This requirement extends to the demand for $\text{spin}^c$-structures on gauge surfaces $S \subset B_3$ in F-theory GUTs, as explained in \cite{BHV09}. Then a choice of $\text{spin}^c$-structure on $N_{C_{\mathbf{R}}|S}$ induces a unique $\text{spin}^c$-structure on $C_{\mathbf{R}}$ as shown in \cite{LM90}.

In general, it is difficult to determine the spin structure on the matter curve $C_{\mathbf{R}}$ needed to analyze the zero mode spectrum. The procedure outlined in \cite{BHV09} assumes that the matter curve $C_{\mathbf{R}}$ belongs to the canonical class of a GUT gauge divisor, but this is not the case for the matter curves in F-theory QSMs \cite{CHLLT19} or the works in \cite{BMPW14, BMW17, Bie18}. In the latter cases, a non-trivial $G_4$-flux was chosen to cancel the ambiguity introduced by the choice of one of the $2^{2g}$ spin bundles on the matter curve $C_{\mathbf{R}}$. To identify the spin bundles in F-theory compactifications such as \cite{BMPW14, BMW17, Bie18, CHLLT19}, it seems necessary to repeat the analysis of \cite{BHV09} for these more general cases, while also investigating the role of 2-torsion F-theory gauge potentials $A_2 \in J^2( \widehat{Y}_4 )$ in the intermediate Jacobian of $\widehat{Y}4$ and their effect on the spin structure of the matter curve $C_{\mathbf{R}}$. However, this is a technically challenging study that has not yet been attempted.

Instead, this note will provide a necessary condition for both the spin bundle and the line bundles induced by the $G_4$-flux, and study all possible combinations of these line bundles to gain a statistical understanding of the likelihood of finding the absence of vector-like exotics in certain representations of the Quadrillion F-theory Standard Models \cite{CHLLT19}.

\section{Necessary root bundle constraints in F-theory} \label{sec:RootBundlesInFTheory}

Let us return to the topic of F-theory gauge potentials and examine them more closely. These potentials are elements of the Deligne cohomology group $H^4_D(\widehat{Y}_4, \mathbb{Z}(2))$. In practical terms, a gauge potential is often defined as $\widehat{\gamma}(\mathcal{A})$ for a ``potential" $\mathcal{A} \in \mathrm{CH}^2(\widehat{Y}_4, \mathbb{Z})$. From the geometry, we can derive a class $\widehat{\gamma}(\mathcal{A}^\prime) \in H^4_D(\widehat{Y}_4, \mathbb{Z}(2))$ and an integer $\xi \in \mathbb{Z}{>0}$ that subject $\mathcal{A}$ to two conditions:
\begin{align}
\gamma( \mathcal{A} ) = G_4 \, , \qquad \xi \cdot \widehat{\gamma}( \mathcal{A} ) \sim \widehat{\gamma} ( \mathcal{A}^\prime ) \, . \label{equ:Conditions}
\end{align}
The statement $\gamma( \mathcal{A} ) = G_4$ implies that $\widehat{\gamma}( \mathcal{A} )$ is a valid F-theory gauge potential for the specific $G_4$-flux. While the $G_4$-flux has a significant effect on the chiral spectrum, the absence of chiral exotics is determined by the second condition in some compactifications. This applies, for example, to the F-theory QSMs \cite{CHLLT19}.

Crucially, the two conditions in equation \cref{equ:Conditions} do not necessarily specify a unique gauge potential $\mathcal{A}$. It is challenging to make assertions about solutions in the Chow group, but in the Deligne cohomology, it is known that if the collection of all $\xi$-th roots of ${\widehat{\gamma}} ( \mathcal{A}^\prime )$ is not empty, it is a coset of the group of all $\xi$-th roots of $0$. Consequently, the number of solutions is $\xi^{2 \cdot \mathrm{dim}_{\mathbb{C}} \left( J^2( \widehat{Y}_4 ) \right)}$. These solutions yield the same chiral spectrum, but they may differ in their vector-like spectrum. This additional flexibility appears to be the primary tool for generating a desirable vector-like spectrum like that of the MSSM.

In many practical scenarios, we can construct $\mathcal{A}^\prime \in \mathrm{CH}^2( \widehat{Y}_4, \mathbb{Z} )$ explicitly. As a result, we have a sufficient level of arithmetic control over $\widehat{\gamma}(\mathcal{A}^\prime)$, and it is reasonable to inquire about the relationship between the induced divisors $D_{\mathbf{R}}( \mathcal{A}^\prime )$ and $D_{\mathbf{R}}( \mathcal{A} )$. They satisfy \cite{BCDLO21}
\begin{align}
\xi \cdot D_{\mathbf{R}} \left( \mathcal{A} \right) \sim D_{\mathbf{R}} \left( \mathcal{A}^\prime \right) \in \mathrm{Pic}( C_{\mathbf{R}} ) \, . \label{equ:Refined}
\end{align}
To clarify, when we consider the F-theory gauge potential $A=\widehat{\gamma}(\mathcal{A})$, it induces a specific divisor $D_{\mathbf{R}}(\mathcal{A})$ such that its $\xi$-th multiple is linearly equivalent to the divisor induced by the F-theory gauge potential $A^\prime =\widehat{\gamma}(\mathcal{A}^\prime)$. Consequently, $D_{\mathbf{R}}(\mathcal{A})$ is a $\xi$-th root of $D_{\mathbf{R}}(\mathcal{A}^\prime)$.

Usually, $\xi$-th roots of $D_{\mathbf{R}} \left( \mathcal{A}^\prime \right)$ do not exist, and even when they do, they are not unique. For the case where $\xi = 2$ and $D_{\mathbf{R}} \left( \mathcal{A}^\prime \right) = K_{{\mathbf{R}}}$, the existence of second roots of the canonical bundle is well-known, and these are the spin structures on $C_{\mathbf{R}}$. If $C_{\mathbf{R}}$ is a smooth curve of genus $g$, then it has $2^{2g}$ spin structures \cite{Ati71, Mum71}. This result can be generalized to $\xi > 2$ and $D_{\mathbf{R}} \left( \mathcal{A}^\prime \right) \neq K_{{\mathbf{R}}}$. Specifically, $\xi$-th roots of $D_{\mathbf{R}} \left( \mathcal{A}^\prime \right)$ exist if and only if $\xi$ divides the degree of $D_{\mathbf{R}} \left( \mathcal{A}^\prime \right)$ and then, there are $\xi^{2g}$ $\xi$-th roots of $D_{\mathbf{R}} \left( \mathcal{A}^\prime \right)$ on a smooth genus $g$ curve \cite{BCDLO21}.

In general, it cannot be assumed that there is a one-to-one correspondence between the $\xi$-th roots of $D_{\mathbf{R}} \left( \mathcal{A}^\prime \right) \in \mathrm{Pic} ( C_{\mathbf{R}})$ and the $\xi$-th roots of $A^\prime = \widehat{\gamma}( \mathcal{A}^\prime ) \in H^4_D( \widehat{Y}_4, \mathbb{Z}(2) )$. Only a subset of the $\xi$-th roots of $D_{\mathbf{R}} \left( \mathcal{A}^\prime \right)$ can be realized from F-theory gauge potentials in $H^4_D( \widehat{Y}_4, \mathbb{Z}(2) )$. Thus, the root bundle constraint in \cref{equ:Refined} is necessary but not sufficient to determine whether the divisor $D_{\mathbf{R}} \left( \mathcal{A} \right)$ arises from an F-theory gauge potential. This is analogous to the challenge of identifying physical spin bundles. The author is unaware of a solution to this problem and will not attempt to provide one in this note. Rather, the goal is to systematically examine all $\xi$-th roots of $D_{\mathbf{R}} \left( \mathcal{A}^\prime \right)$ and all spin bundles on $C_{\mathbf{R}}$ in order to identify combinations of root and spin bundles that result in a line bundle whose cohomologies satisfy the physical demand for the presence or absence of vector-like pairs. This study is a first step towards understanding the likelihood of finding solutions to 4-dimensional F-theory compactifications without vector-like exotics.

As an example, look at the F-theory QSMs \cite{CHLLT19}. These geometries are defined in terms of toric 3-folds with $\overline{K}^3_{B_3} \in \{ 6, 10, 18, 30 \}$. The matter curves are defined by sections $s_i \in H^0 \left( B_3, \overline{K}_{B_3} \right)$. An explicit study finds the following root bundle constraints \cite{BCDLO21}:
\begin{align}
\begin{tabular}{cc}
\toprule
curve & root bundle constraint \\
\midrule
$C_{(\mathbf{3},\mathbf{2})_{1/6}} = V( s_3, s_9 )$ & $P_{(\mathbf{3},\mathbf{2})_{1/6}}^{\otimes 2 \overline{K}^3_{B_3}} = K_{{(\mathbf{3},\mathbf{2})_{1/6}}}^{\otimes \left( 6 + \overline{K}^3_{B_3} \right)}$ \\
$C_{(\mathbf{1},\mathbf{2})_{-1/2}}= V \left( s_3, P_H \right)$ & $P_{(\mathbf{1},\mathbf{2})_{-1/2}}^{\otimes 2 \overline{K}^3_{B_3}} = K_{{(\mathbf{1},\mathbf{2})_{-1/2}}}^{\otimes \left( 4 + \overline{K}^3_{B_3} \right)} \otimes \mathcal{O}_{(\mathbf{1},\mathbf{2})_{-1/2}} \left( - 30 Y_1 \right)$ \\
$C_{(\overline{\mathbf{3}},\mathbf{1})_{-2/3}} = V( s_5, s_9 )$ & $P_{(\overline{\mathbf{3}},\mathbf{1})_{-2/3}}^{\otimes 2 \overline{K}^3_{B_3}} = K_{{(\overline{\mathbf{3}},\mathbf{1})_{-2/3}}}^{\otimes \left( 6 + \overline{K}^3_{B_3} \right)}$ \\
$C_{(\overline{\mathbf{3}},\mathbf{1})_{1/3}} = V \left( s_9, P_R \right)$ & $P_{(\overline{\mathbf{3}},\mathbf{1})_{1/3}}^{\otimes 2 \overline{K}^3_{B_3}} = K_{{(\overline{\mathbf{3}},\mathbf{1})_{1/3}}}^{\otimes \left( 4 + \overline{K}^3_{B_3} \right)} \otimes \mathcal{O}_{(\overline{\mathbf{3}},\mathbf{1})_{1/3}} \left( - 30 Y_3 \right)$ \\
$C_{(\mathbf{1},\mathbf{1})_{1}} = V( s_1, s_5 )$ & $P_{(\mathbf{1},\mathbf{1})_{1}}^{\otimes 2 \overline{K}^3_{B_3}} = K_{{(\mathbf{1},\mathbf{1})_{1}}}^{\otimes \left( 6 + \overline{K}^3_{B_3} \right)}$ \\
\bottomrule
\end{tabular} \label{tab:RootBundles}
\end{align}
In this table, we make use of two polynomials: $P_H = s_2 s_5^2 + s_1 ( s_1 s_9 - s_5 s_6 )$ and $P_R = s_3 s_5^2 + s_6 ( s_1 s_6 - s_2 s_5 )$. It is important to note that the line bundles on both the Higgs curve $C_{(\mathbf{1},\mathbf{2}){-1/2}}$ and the curve $C{(\overline{\mathbf{3}},\mathbf{1})_{1/3}}$ are dependent on the Yukawa points $Y_1 = V( s_3, s_5, s_9 )$ and $Y_3 = V( s_3, s_6, s_9 )$, respectively. Furthermore, it is important to keep in mind that if two divisors $D$ and $E$ are linearly equivalent, i.e., $D \sim E$, then $n \cdot D \sim n \cdot E$ for any integer $n$. However, the converse is not true, and that is why we do not cancel common factors.

\section{Vector-like spectra of F-theory QSMs from limit roots} \label{sec:VLSpectraFromLimitRoots}

To construct root bundles that meet the constraints in \cref{tab:RootBundles} and have the phenomenologically desired cohomologies, we begin by deforming the curve $C_{\mathbf{R}}$ to a nodal curve $C^{\bullet}_{\mathbf{R}}$. Deforming to a nodal curve has two benefits. Firstly, constructing root bundles on smooth curves is difficult, while constructing them on nodal curves via so-called limit roots has been thoroughly detailed and involves combinatorial tasks \cite{CCC07}. Secondly, for F-theory QSMs, the nodal curve is invariant across a range of different geometries, so we can approximate the vector-like spectra for many F-theory QSMs with a single computation. To fully appreciate these advantages, we will provide a summary of limit root bundles in \cref{sec:LimitRoots} and explain the invariance of the nodal matter curve in \cref{sec:FRST-Invariance}.

Certainly, we eventually want to determine the cohomologies of root bundles on smooth, irreducible matter curves, so we study how those cohomologies relate to the cohomologies of limit roots. The pushforward of limit roots $P^\circ_{\mathbf{R}}$ through the blow-up map $\pi \colon C^\circ_{\mathbf{R}} \to C^\bullet_{\mathbf{R}}$ keeps the number of global sections the same, that is, $h^0( C^\circ_{\mathbf{R}}, P^\circ_{\mathbf{R}} ) = h^0( C^\bullet_{\mathbf{R}}, P^\bullet_{\mathbf{R}} )$. In principle, we can trace these roots $P^\bullet_{\mathbf{R}}$ along the deformation $C^\bullet_{\mathbf{R}} \to C_{\mathbf{R}}$ to find roots $P_{\mathbf{R}}$ on the original curve $C_{\mathbf{R}}$. However, this deformation can alter the number of sections. For the deformation $C^\bullet_{\mathbf{R}} \to C_{\mathbf{R}}$, where a nodal curve is transformed into a smooth, irreducible curve, it is known that the number of sections is an upper semi-continuous function. Thus, the number of sections either stays the same or decreases:
\begin{align}
h^0( C_{\mathbf{R}}, P_{\mathbf{R}} ) \leq h^0( C^{\bullet}_{\mathbf{R}}, P^{\bullet}_{\mathbf{R}} ) = \chi( P_{\mathbf{R}} ) + \delta \, , \qquad \delta \in \mathbb{Z}_{\geq 0} \, .
\end{align}
In order to find roots for which the number of sections remains constant upon deformation to the original curve, we can look for instances where equality holds. One such instance is the generic case $\delta=0$, in which $h^0( C^{\bullet}_{\mathbf{R}}, P^{\bullet}_{\mathbf{R}} ) = \chi( P_{\mathbf{R}} )$. This is because the number of sections is already minimal on $C^{\bullet}_{\mathbf{R}}$ and therefore must remain constant throughout the deformation. As a result, the number of limit roots on the partial blow-ups $C^\circ_{\mathbf{R}}$ of the nodal curve $C^\bullet_{\mathbf{R}}$ that have the generic number of global sections provides a lower bound on the number of roots $P_{\mathbf{R}}$ without vector-like exotics.

For 33 F-theory QSMs families, which we select in \cref{subsec:SelectedQSMs}, the predictive powers of these counts are strongest. This is due to the limitation that it is not yet clear which of the mathematical root bundles stem from F-theory gauge potentials. In \cref{subsec:TowardsBNOfLimitRoots}, we then outline techniques to count the number of limit roots with prescribed number of global sections. This allows us to gauge -- statistically speaking -- how likely absence of vector-like exotics is realized within the F-theory QSMs. The computer implementation used for to obtain these counts cannot (yet) determine the number of global sections for all limit roots. Rather, if we want to see the complete picture, a few remaining cases must be treated by hand. For the QSM familiy $B_3( \Delta_4^\circ )$, this is outlined in \cref{subsec:D4}. We find that more than $99.995\%$ of the limit roots admit exactly three global sections, just as desired for string phenomenology.

\subsection{Limit roots} \label{sec:LimitRoots}

To recap, on a smooth curve $C$, $n$-th roots $P$ of a line bundle $L$ exist when $\deg(L)$ is divisible by $n$. However, this is not the case for reducible, nodal curves $C^\bullet$. In this situation, a root $P^\bullet$ of a line bundle $L^\bullet$ on such curves should restrict to a root on the irreducible components, but even if $n$ divides the degree of $L^\bullet$, it may not divide $\deg(L|_Z)$ for some irreducible component $Z$ of $C^\bullet$. This issue is elegantly avoided by utilizing limit $n$-th root bundles $P^\circ$ on (partial) blow-ups $C^\circ$ of $C^\bullet$, as originally introduced in \cite{CCC07}. These limit $n$-th root bundles $P^\circ$ are determined by weighted graphs, which can be exploited to make the task of section-counting manageable. In the following section, we will briefly outline this combinatorial approach to limit roots. Those interested in further information on limit roots may consult \cite{Jar01, Jar98}.

To represent a connected nodal curve $C^\bullet$ with arithmetic genus $g$, we create a \emph{dual graph} $\Pi_{C^\bullet}$. Each vertex of $\Pi_{C^\bullet}$ corresponds to an irreducible component $C^\bullet_i$ of $C^\bullet$. Each half-edge extending from a vertex $C^\bullet_i$ is a node on $C^\bullet_i$. If a node is present on both $C^\bullet_i$ and $C^\bullet_j$, the half-edges connected to $C^\bullet_i$ and $C^\bullet_j$ come together to form an edge.

\begin{example}
We can represent the dual graph $\Pi_{E^\bullet}$ of the nodal curve $E^\bullet$ with two rational components $C_1$ and $C_2$ intersecting at two nodes as follows:
\begin{equation}
\begin{tikzpicture}[baseline=(current  bounding  box.center)]
\def\s{1};
\path[-, out = 45, in = 135, looseness = 0.8] (-\s,0) edge (\s,0);
\path[-, out = -45, in = -135, looseness = 0.8] (-\s,0) edge (\s,0);
\node at (-\s,0) [circle, draw, fill=white, label=left:$C_1$] {};
\node at (\s,0)  [circle, draw, fill=white, label=right:$C_2$] {};
\end{tikzpicture}
\end{equation}
\end{example}

Suppose we have a nodal curve $C^\bullet$ and a blow-up $\pi: C^\circ \rightarrow C^\bullet$. If $n_i$ is a node on $C^\bullet$, then the exceptional components over $n_i$ are denoted by $\mathcal{E}_i$, each of which is isomorphic to $\mathbb{P}^1$. Let $C^N$ be the complement of the exceptional components in $C^\circ$. Then, the map $\pi|_{C^N}: C^N \rightarrow C^\bullet$ is the normalization of $C^\bullet$. The points in $(\pi|_{C^N})^{-1}(n_i)$ are called the exceptional nodes and consist of the intersection of the exceptional components with $C^N$. A \emph{full blow-up} of $C^\bullet$ is a blow-up at all the nodes, while a \emph{partial blow-up} is a blow-up at a proper subset of the nodes.

To begin with, consider a positive integer $n$ and a line bundle $L^\bullet$ on the nodal curve $C^\bullet$ such that $n$ divides the degree of $L^\bullet$. Let $\Delta_{C^\bullet}$ be a subset of the nodes of $C^\bullet$, which may or may not be the entire set.

\begin{definition}
A \emph{limit $n$-th root} of $L^\bullet$ associated to $\Delta_{C^\bullet}$ is a triple $(C^\circ, P^\circ, \alpha)$ consisting of: 
\begin{itemize}
\item the (partial) blow-up $\pi: C^\circ \rightarrow C^\bullet$,
\item a line bundle $P^\circ$ on $C^\circ$,
\item a homomorphism $\alpha: (P^\circ)^n \rightarrow \pi^*(L^\bullet)$,
\end{itemize}
satisfying the following properties:
\begin{enumerate}
\item $\deg(P^\circ|_{\mathcal{E}_i}) = 1$ for every exceptional component $\mathcal{E}_i$,
\item $\alpha$ is an isomorphism at all points of $C^\circ$ except for the exceptional components,
\item for every exceptional component $\mathcal{E}_i$ of $C^\circ$, the sum of the orders of vanishing of $\alpha$ at the exceptional nodes $p_i$ and $q_i$ is equal to $n$. 
\end{enumerate}
\end{definition}

The combinatorial information of limit $n$-th roots over $C^\bullet$ is represented by weighted graphs, which are based on the dual graph of the nodal curve $C^\bullet$. Conversely, these graphs allow for the construction and retrieval of limit $n$-th roots. Although there is not a one-to-one correspondence between the weighted graphs and limit roots, these graphs provide a useful way to parametrize the limit roots. To begin, let us define the weighted graphs in question. The set of exceptional nodes corresponding to $\Delta_{C^\bullet}$ is denoted by $\widetilde{\Delta}_{C^\bullet}$.

\begin{definition}\label{Defi:weightedgraph}
The \emph{weighted graph linked to a limit $n$-th root $(C^\circ, P^\circ, \alpha)$ of $L^\bullet$} refers to the dual graph $\Pi_{C^\bullet}$ with weights assigned using a weight function
\begin{align}
w: \widetilde{\Delta}_{C^\bullet} \rightarrow {1, \dots, n-1} \, ,
\end{align}
where $w(p_i) = u_i$ and $w(q_i) = v_i$ denote the orders at which $\alpha$ vanishes at $p_i$ and $q_i$ respectively.
\end{definition}

These weighted graphs have two inherent characteristics:
\begin{itemize}
\item[(A)] The sum of weights assigned to vertices $p_i$ and $q_i$ equals $u_i+v_i$, which in turn is equal to $n$.
\item[(B)] For each irreducible component $C^\bullet_i$ of $C^\bullet$, the sum of weights assigned to the vertex corresponding to $C^\bullet_i$ is congruent to $\deg_{C^\bullet_i} L^\bullet$ modulo $n$.
\end{itemize}

\begin{example}
 We return to the nodal curve $E^\bullet$ and wish to find the limit 2nd roots of $K_{E^\bullet}$. If $C^\bullet_i$ is a component of $E^\bullet$, then set $k_i = \# C^\bullet_i \cap (\overline{H^\bullet \setminus C^\bullet_i})$. Therefore, \mbox{$\deg(K_{E^\bullet}|_{C^\bullet_i}) = 2g(C^\bullet_i) -2 + k_i = -2 + k_i$}, and the multi-degree of $K_{E^\bullet}$ is
\begin{align}
\left( \deg(K_{E^\bullet}|_{C_1} ), \deg(K_{H^\bullet}|_{C_2} \right) = (0, 0) \, ,
\end{align}
which has total degree is $0$. Here are the descriptions of the two weighted graphs connected to the (partial) blow-up limit 2nd roots of $K_{E^\bullet}$:
\begin{equation}
\begin{tikzpicture}[baseline=(current  bounding  box.center)]
\def\s{1};
\def\d{5};
\path[-, every node/.append style={fill=white}, out = 45, in = 135, looseness = 0.8] (-\s,0) edge node[pos=0.35] {1} node[pos=0.65] {1} (\s,0);
\path[-, every node/.append style={fill=white}, out = -45, in = -135, looseness = 0.8] (-\s,0) edge node[pos=0.35] {1} node[pos=0.65] {1} (\s,0);
\node at (-\s,0) [circle, draw, fill=white, label=left:$C_1$] {$0$};
\node at (\s,0)  [circle, draw, fill=white, label=right:$C_2$] {$0$};
\path[-, out = 45, in = 135, looseness = 0.8] (-\s+\d,0) edge (\s+\d,0);
\path[-, out = -45, in = -135, looseness = 0.8] (-\s+\d,0) edge (\s+\d,0);
\node at (-\s+\d,0) [circle, draw, fill=white, label=left:$C_1$] {$0$};
\node at (\s+\d,0)  [circle, draw, fill=white, label=right:$C_2$] {$0$};
\end{tikzpicture}
\end{equation}
\end{example}
The labels inside the vertices are the multi-degrees of $K_{E^\bullet}$, while the labels outside the vertices are the weights.

It is important to acknowledge that any weighted graph with an underlying graph of $\Pi_{C^\bullet}$ and a weight function $w: \widetilde{\Delta}_{C^\bullet} \rightarrow {1, \dots, n-1}$ that satisfies the conditions (A) and (B) contains information about a limit $n$-th root $(C^\circ, P^\circ, \alpha)$ of $L$. Indeed, this specific weighted graph is the same as the weighted graph linked to $(C^\circ, P^\circ, \alpha)$ of $L^\bullet$.

For any nodal curve $C^\bullet$ and line bundle $L^\bullet$ on $C^\bullet$, there exists a total of $n^{b_1(\Pi_{C^\bullet})}$ weighted subgraphs that fulfill conditions (A) and (B), where
\begin{align}
b_1(\Pi_{C^\bullet}) = \# \text{edges} + \#\text{connected components} - \# \text{vertices} , ,
\end{align}
represents the first Betti number of $\Pi_{C^\bullet}$. It is important to note that this includes all of the weighted subgraphs whose edge sets correspond to subsets of $\Delta_{C^\bullet}$.

The relationship between limit $n$-th roots and weighted graphs that fulfill conditions $(A)$ and $(B)$ is not a one-to-one mapping. In fact, creating a limit root involves selecting a root $P^N$ of $(\pi|_{C^N})^*(L^\bullet)(-\sum (u_ip_i + v_i q_i))$. By doing a precise calculation, it can be determined that there are $n^{2g}$ limit $n$-th roots \cite{CCC07}.

\begin{example}
We describe the full blow-up limit 2nd roots of $K_{E^\bullet}$:
\begin{enumerate}
 \item Blow-up at the set of all nodal singularities, and denote the exceptional component at the $i$-th node by $\mathcal{E}_i \cong \mathbb{P}^1$. This $\mathbb{P}^1$ touches $E_i$ at the exceptional node $p_i$ and $\Gamma$ at $q_i$.
 \item Let $E^N$ be the (full) normalization of $E^\bullet$, and consider the bundle with multi-degree $(-2, -2)$. This bundle admits exactly one root and this root has multi-degree $\left( -1, -1 \right)$.
 \item Pick the 2nd root $P^N$, and glue to it a degree one bundle over every $\mathcal{E}_i$. The resulting \emph{limit 2nd root $P^\circ$ of $K_{E^\bullet}$} can be represented as follows:
\begin{equation}
\begin{tikzpicture}[baseline=(current  bounding  box.center)]
\def\s{2};
\path[-, out = 45, in = 180, looseness = 0.8] (-\s,0) edge (0,0.5\s);
\path[-, out = -45, in = -180, looseness = 0.8] (-\s,0) edge (0, -0.5\s);
\path[-, out = 135, in = 0, looseness = 0.8] (\s,0) edge (0,0.5\s);
\path[-, out = -135, in = 0, looseness = 0.8] (\s,0) edge (0, -0.5\s);

\node at (-\s,0) [circle, draw, fill=white, label=left:$C_1$] {$-1$};
\node at (\s,0)  [circle, draw, fill=white, label=right:$C_2$] {$-1$};
\node at (0,0.5\s) [circle, fill=white] {$1$};
\node at (0,0.5\s) [circle, draw, pattern=dots, pattern color = black!70, label={[label distance=0.01cm]-10:$\mathcal{E}_1$}] {$1$};
\node at (0,-0.5\s) [circle, fill=white] {$1$};
\node at (0,-0.5\s)  [circle, draw, pattern=dots, pattern color = black!70, label={[label distance=0.01cm]170:$\mathcal{E}_2$}] {$1$};

\end{tikzpicture}
\label{equ:Graph1}
\end{equation}
\end{enumerate}
We obtain a distinct limit 2nd root of $K_{E^\bullet}$ by not performing any node blow-ups:
\begin{equation}
\begin{tikzpicture}[baseline=(current  bounding  box.center)]
\def\s{1.0};
\path[-, out = 45, in = 135, looseness = 0.8] (-\s,0) edge (\s,0);
\path[-, out = -45, in = -135, looseness = 0.8] (-\s,0) edge (\s,0);
\node at (-\s,0) [circle, draw, fill=white, label=left:$C_1$] {$0$};
\node at (\s,0)  [circle, draw, fill=white, label=right:$C_2$] {$0$};
\end{tikzpicture}
\label{equ:Graph2}
\end{equation}
\end{example}

On a smooth curve with genus $g=1$, there exist four different 2nd roots of $K_C$, which seems to be in contrast to the case of a nodal curve with genus $g=1$ for which we only found two distinct weighted subgraphs. To obtain $r^{2g}$ limit $r$th roots, we note that each weighted graph accounts for multiple limit roots. This multiplicity is known as the \emph{geometric multiplicity} \cite{CCC07}. The number of distinct limit roots associated to a weighted subgraph $\Delta^w$ is given by $\mu = r^{2g^\nu + b_1(\Gamma_C)}$, where $g^\nu$ is the arithmetic genus of the normalization \cite{BCDO22}.

\begin{example}
In the case of our example of a nodal $g = 1$ curve, we have $g^\nu = 0$, which means that the geometric multiplicity is $\mu = r^{2g^\nu + b_1(\Gamma_C)} = r^{b_1(\Gamma_C)} = 2$. This implies that the weighted graphs in \cref{equ:Graph1} and \cref{equ:Graph2} each encode two distinct limit root bundles, and thus we obtain a total of $r^{2g} = 4$ limit spin bundles, as expected.
\end{example}

\subsection{FRST-invariant graphs} \label{sec:FRST-Invariance}

The F-theory QSM geometries arise from desingularizations of toric K3-surfaces. Such desingularizations were first studied in \cite{Bat94} and correspond to 3-dimensional, reflexive lattice polytopes $\Delta \subset M_{\mathbf{R}}$ and their polar duals $\Delta^\circ \subset N_{\mathbf{R}}$ defined by $\left\langle \Delta, \Delta^\circ \right\rangle \geq -1$. The complete list of such 3-dimensional polytopes is available in \cite{KS98}. We follow \cite{BCL21} and denote the i-th polytope in this list as $\Delta_i^\circ \subset N_{\mathbf{R}}$.

Given a lattice polytope $\Delta \subset M_{\mathbf{R}}$, one can derive its normal fan $\Sigma_\Delta$ by using the facet normals of $\Delta$ as ray generators. The vertices of $\Delta$ correspond to the maximal cones of $\Sigma_\Delta$. Although a CY-hypersurface in the toric variety $X_{\Delta} \equiv X_{\Sigma_{\Delta}}$ may not be smooth, its singularities can be resolved through a process known as a \emph{maximal projective crepant partial} (MPCP) desingularization, as discussed in \cite{Bat94}. The aforementioned resolutions are sometimes referred to as maximal projective subdivisions of the normal fan \cite{CK99}. An MPCP desingularization corresponds to a fine, regular, star triangulation (FRST) $\Sigma(T)$ of the lattice polytope's polar dual, $\Delta^\circ$. In this context, \emph{star} means that every simplex in the triangulation contains the origin, \emph{fine} indicates that every lattice point in $\Delta^\circ$ is used as a ray generator\footnote{The described property is also known as being \emph{full}\cite{DRS10}.}, and \emph{regular} implies that $X_{\Sigma(T)}$ is a projective variety. Consequently, $\Sigma(T)$ defines a \emph{maximal} projective refinement of the normal fan $\Sigma_{\Delta}$. In applications involving toric K3-surfaces, the resulting $X_{\Sigma(T)}$ is smooth \cite{BCL21}, due to results in \cite{Bat94, CK99, CLS11}.

The F-theory QSM geometries have a requirement that $\overline{K}_{X_{\Sigma(T)}}^3$ must be one of the values ${6, 10, 18, 30}$. Out of the $4319$ polytopes listed in \cite{KS98}, only $708$ satisfy this condition \cite{CHLLT19}. These polytopes have multiple FRSTs, with an approximate count provided in \cite{HT17}. The family of toric 3-folds obtained from the various FRSTs of $\Delta^\circ$ is denoted by $B_3(\Delta^\circ)$. For each F-theory QSM, a nodal quark-doublet curve can be introduced \cite{BCDLO21}
\begin{align}
C_{(\mathbf{3},\mathbf{2})_{1/6}}^\bullet = \bigcup \limits_{i \in L}{V( x_i, s_9 )} \, ,
\label{equ:NodalQuarkDoubletCurve}
\end{align}
where $L$ is the collection of all lattice points of $\Delta^\circ$ and $V( s_9 )$ is the K3-surface. 

We make the assumption that the element $s_9 \in H^0(X_{\Sigma(T)}, \overline{K}_{X_{\Sigma(T)}})$ is generic. In \cite{BCL21}, it was argued that the set $V(x_i, s_9)$ is either empty, consists of a single irreducible component, or is a finite collection of smooth $\mathbb{P}^1$s (further background on this topic can be found in \cite{Bat94, PS97, CK99, Roh04, Kre10}). The number of intersection points among the irreducible components of $C_{(\mathbf{3}, \mathbf{2})_{1/6}}^\bullet$ is equal to their topological intersection number, which in turn depends only on $\Delta^\circ$ \cite{BCL21}. Therefore, the dual graph of $C_{(\mathbf{3}, \mathbf{2})_{1/6}}^\bullet$ is the same for all FRSTs of $\Delta^\circ$, meaning that it is identical for all spaces in $B_3(\Delta^\circ)$. This is an impressive finding! Despite there being an enormous number of F-theory QSM geometries in $B_3(\Delta^\circ_8)$ (around $10^{15}$), it is possible to estimate the vector-like spectra of all of them by counting the global sections of limit roots on a single nodal curve.

\subsection{Selection among the F-theory QSMs} \label{subsec:SelectedQSMs}

To find the number of limit roots $P_{(\mathbf{3},\mathbf{2})_{1/6}}^\bullet$ that satisfy the conditions of having three global sections, we need to investigate the $(6 + \overline{K}^3_{B_3})$-th power of the canonical bundle on $C_{(\mathbf{3},\mathbf{2}){1/6}}^\bullet$ (cf. \cref{tab:RootBundles}). There are $(2 \overline{K}^3_{B_3})^{2g}$ such roots on a genus $g$ curve. In comparison, there are $(2\overline{K}^3_{B_3})^{2h^{2,1}(\widehat{Y}_4)}$ inequivalent F-theory gauge potentials, but we do not yet know which of these roots are physical. Therefore, we focus on F-theory QSMs that have at least as many F-theory gauge potentials as roots on $C_{(\mathbf{3},\mathbf{2})_{1/6}}$, which means that many of the roots can be expected to come from F-theory gauge potentials. There are 37 F-theory QSM families that satisfy $h^{2,1}(\widehat{Y}_4) \geq g$, but four of these families have $C^\bullet_{(\mathbf{3},\mathbf{2})_{1/6}}$ with a component whose genus is greater than one. Since this makes it difficult to compute the number of global sections, we ignore these four QSM families for now. The triangulations of the remaining 33 polytopes provide most of the $\mathcal{O}(10^{15})$ F-theory QSMs. \cite{HT17,CHLLT19}.

\subsection{First steps towards Brill-Noether theory of limit roots} \label{subsec:TowardsBNOfLimitRoots}

We will demonstrate how to use section counting techniques to analyze limit roots in F-theory QSMs \cite{CHLLT19}, by using the example of the fourth polytope $\Delta_4^\circ$ in the Kreuzer-Skarke database \cite{KS98}. We examine the dual graph of $C^\bullet_{(\mathbf{3}, \mathbf{2})_{1/6}}$ for the $\mathcal{O}( 10^{11} )$ geometries $B_3( \Delta_4^\circ )$ \cite{HT17}, which can be represented as follows:
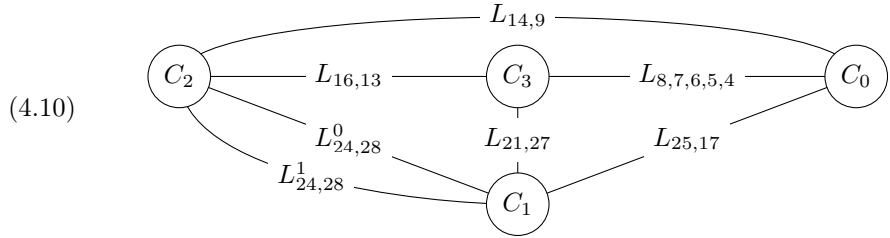
\begin{equation}
\begin{tikzpicture}[baseline=(current  bounding  box.center)]
      
      % define parameters
      \def\s{4.5};
      \def\h{1.7};
      
      % draw paths between curves
      \path[-] (-\s,0) edge node[fill = white] {$L_{16,13}$} (0,0);
      \path[-] (-\s,0) edge node[fill = white] {$L^0_{24,28}$} (0,-\h);
      \path[-] (\s,0) edge node[fill = white] {$L_{8,7,6,5,4}$} (0,0);
      \path[-] (\s,0) edge node[fill = white] {$L_{25,17}$} (0,-\h);
      \path[-] (0,0) edge node[fill = white] {$L_{21,27}$} (0,-\h);
      \path[-, out = -90, in = 180, looseness = 0.7] (-\s,0) edge node[fill = white] {$L^1_{24,28}$} (0,-\h);
      \path[-, out = 90, in = 90, looseness = 0.3] (-\s,0) edge node[fill = white] {$L_{14,9}$} (\s,0);
      
      % draw curves
      \node at (1*\s,0) [circle, draw, fill=white]{$C_0$};
      \node at (0,-\h) [circle, draw, fill=white]{$C_1$};
      \node at (-1*\s,0) [circle, draw, fill=white]{$C_2$};
      \node at (0,0) [circle, draw, fill=white]{$C_3$};
      
\end{tikzpicture}
\label{equ:Delta4FullDualGraph}
\end{equation}
In this diagram, the edges indicate nodal singularities, while the vertices represent the irreducible components of $C^\bullet_{(\mathbf{3}, \mathbf{2}){1/6}}$. The four curves $C_i$ are all smooth $\mathbb{P}^1$s. Additionally, $L_{i_1,...,i_k}$ denotes a link of $k$ smooth, genus-zero curves $C_{i_j}$, such that each pair of adjacent curves intersects at exactly one node. Therefore, as a dual graph, $L_{i_1, ..., i_k}$ corresponds to a link of $k$ vertices, with the vertices ordered such that $C_{i_1}$ intersects the $\mathbb{P}^1$ on the left and $C_{i_k}$ intersects the $\mathbb{P}^1$ on the right. For example, in the case of $L_{21, 27}$, $C_{21}$ intersects $C_3$, while $C_{27}$ intersects $C_1$.

Remember that for a curve $C$ with dual graph $\Gamma_C$, there are $r^{b_1(\Gamma_C)}$ weighted subgraphs that meet the conditions (C1) and (C2) in \cite{CCC07}. In the case where $C$ is a link of rational curves, we have $b_1(\Gamma_C) = 0$, and therefore there is only one possible weighted subgraph, which corresponds to the full blow-up. As a result, we can substitute the dual graph in \cref{equ:Delta4FullDualGraph} with the following graph, which simplifies the process of counting limit root bundles \cite{BCL21, BCDO22}:
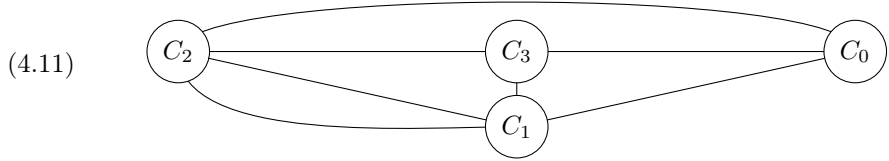
\begin{equation}
\begin{tikzpicture}[baseline=(current  bounding  box.center)]
      \def\s{4.5};
      \def\h{1.0};
      
      \path[-] (-\s,0) edge (0,0);
      \path[-] (-\s,0) edge (0,-\h);
      \path[-] (\s,0) edge (0,0);
      \path[-] (\s,0) edge (0,-\h);
      \path[-] (0,0) edge (0,-\h);
      \path[-, out = -90, in = 180, looseness = 0.7] (-\s,0) edge (0,-\h);
      \path[-, out = 90, in = 90, looseness = 0.25] (-\s,0) edge (\s,0);
      
      \node at (1*\s,0) [circle, draw, fill=white]{$C_0$};
      \node at (0,-\h) [circle, draw, fill=white]{$C_1$};
      \node at (-1*\s,0) [circle, draw, fill=white]{$C_2$};
      \node at (0,0) [circle, draw, fill=white]{$C_3$};
\end{tikzpicture}
\end{equation}
The first counts of limit roots with exactly three global sections were presented in \cite{BCL21}. The focus of this study was on full blow-up limit roots, which means that each nodal singularity is substituted by an exceptional $\mathbb{P}^1$. In the case of $\Delta_4^\circ$, the dual graph of this curve $C^\circ_{(\mathbf{3}, \mathbf{2})_{1/6}}$ takes the following shape:
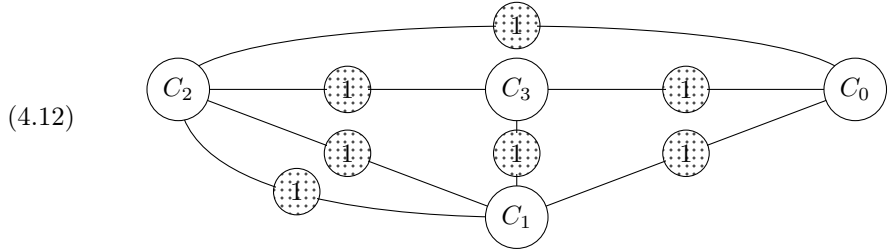
\begin{equation}
\begin{tikzpicture}[baseline=(current  bounding  box.center)]
      \def\s{4.5};
      \def\h{1.7};
      
      \path[-] (-\s,0) edge (0,0);
      \path[-] (-\s,0) edge (0,-\h);
      \path[-] (\s,0) edge (0,0);
      \path[-] (\s,0) edge (0,-\h);
      \path[-] (0,0) edge (0,-\h);
      \path[-, out = -90, in = 180, looseness = 0.8] (-\s,0) edge (0,-\h);
      \path[-, out = 90, in = 90, looseness = 0.32] (-\s,0) edge (\s,0);
      
      \node at (1*\s,0) [circle, draw, fill=white]{$C_0$};
      \node at (0,-\h) [circle, draw, fill=white]{$C_1$};
      \node at (-1*\s,0) [circle, draw, fill=white]{$C_2$};
      \node at (0,0) [circle, draw, fill=white]{$C_3$};
      
      \node at (0.5*\s,0) [fill=white, scale = 2.3] {};
      \node at (0.5*\s,0) [circle, draw, pattern=dots, pattern color = black!70]{$1$};
      \node at (0.5*\s,-0.5*\h) [fill=white, scale = 2.3] {};
      \node at (0.5*\s,-0.5*\h) [circle, draw, pattern=dots, pattern color = black!70]{$1$};
      \node at (-0.5*\s,0) [fill=white, scale = 2.3] {};
      \node at (-0.5*\s,0) [circle, draw, pattern=dots, pattern color = black!70]{$1$};
      \node at (-0.5*\s,-0.5*\h) [fill=white, scale = 2.3] {};
      \node at (-0.5*\s,-0.5*\h) [circle, draw, pattern=dots, pattern color = black!70]{$1$};
      \node at (0*\s,-0.5*\h) [fill=white, scale = 2.3] {};
      \node at (0*\s,-0.5*\h) [circle, draw, pattern=dots, pattern color = black!70]{$1$};
      \node at (0*\s,0.5*\h) [fill=white, scale = 2.3] {};
      \node at (0*\s,0.5*\h) [circle, draw, pattern=dots, pattern color = black!70]{$1$};
      \node at (-0.65*\s,-0.8*\h) [fill=white, scale = 2.3] {};      
      \node at (-0.65*\s,-0.8*\h) [circle, draw, pattern=dots, pattern color = black!70]{$1$};
      
\end{tikzpicture}
\end{equation}
In this diagram, the exceptional $\mathbb{P}^1$s are marked with grey dots in the background. It's worth noting that for each exceptional $\mathbb{P}^1$, $\left. P \right|_{E_i} \cong \mathcal{O}_{\mathbb{P}^1}(1)$, as established in \cite{CCC07}. The (local) sections on the exceptional $\mathbb{P}^1$s are uniquely determined by the need to connect the sections on the components $C_i$ of $C^\bullet_{(\mathbf{3}, \mathbf{2})_{1/6}}$, which is why they were referred to as ``bridging sections'' in \cite{BCDLO21}. Consequently, the number of global sections of $P^\bullet$ is the same as that of the restriction of $P^\bullet$ to the following completely disconnected curve:
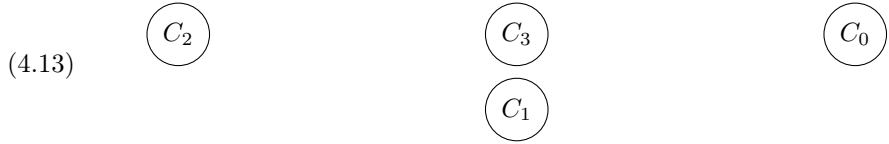
\begin{equation}
\begin{tikzpicture}[baseline=(current  bounding  box.center)]
      \def\s{4.5};
      \def\h{1.0};
      \node at (1*\s,0) [circle, draw, fill=white]{$C_0$};
      \node at (0,-\h) [circle, draw, fill=white]{$C_1$};
      \node at (-1*\s,0) [circle, draw, fill=white]{$C_2$};
      \node at (0,0) [circle, draw, fill=white]{$C_3$};
\end{tikzpicture}
\end{equation}
Consequently, it holds
\begin{equation}
h^0(C^\circ_{\textbf{R}}, P^\circ_{\textbf{R}}) = \sum_Z h^0(Z, P^\circ_{\textbf{R}}|_{Z}) \, ,
\end{equation}
where $Z$ is an irreducible component of $C^\bullet_{\textbf{R}}$. A computational analysis of the 33 chosen F-theory QSMs yielded the following counts of limit roots possessing precisely three global sections \cite{BCL21, BCDO22}:
\begin{align}
\begin{tabular}{c|c|ccc|c}
\toprule
$\overline{K}^3_{B_3}$ & Polytope & $\check{N}^{(3)}_P$ & $\mu$ & $N_{\text{total}}$ & $\frac{\check{N}^{(3)}_P \cdot \mu}{N_{\text{total}}}$ [\%]\\
\midrule
\multirow{4}{*}{$6$} & $\Delta^\circ_{8}$ & $142560$ & $12^3$ & $12^8$ & $57.3$ \\
& $\Delta^\circ_{4}$ & $11110$ & $12^4$ & $12^8$ & $53.6$ \\
& $\Delta^\circ_{134}$ & $10010$ & $12^4$ & $12^8$ & $48.7$ \\
& $\Delta^\circ_{128}$, $\Delta^\circ_{130}$, $\Delta^\circ_{136}$, $\Delta^\circ_{236}$ & $8910$ & $12^4$ & $12^8$ & $42.0$ \\
\midrule
\multirow{21}{*}{$10$} & $\Delta^\circ_{88}$ & $781.680.888$ & $20^5$ & $20^{12}$ & $61.1$ \\ 
& $\Delta^\circ_{110}$ & $738.662.983$ & $20^5$ & $20^{12}$ & $57.8$ \\
& $\Delta^\circ_{272}$, $\Delta^\circ_{274}$ & $736.011.640$ & $20^5$ & $20^{12}$ & $57.5$ \\
& $\Delta^\circ_{387}$ & $733.798.300$ & $20^5$ & $20^{12}$ & $57.3$ \\
& $\Delta^\circ_{798}$, $\Delta^\circ_{808}$, $\Delta^\circ_{810}$, $\Delta^\circ_{812}$ & $690.950.608$ & $20^5$ & $20^{12}$ & $54.0$ \\
\cmidrule{2-6}
& $\Delta^\circ_{254}$ & $35.004.914$ & $20^6$ & $20^{12}$ & $54.7$ \\
& $\Delta^\circ_{302}$ & $34.908.682$ & $20^6$ & $20^{12}$ & $54.7$ \\
& $\Delta^\circ_{52}$ & $34.980.351$ & $20^6$ & $20^{12}$ & $54.7$ \\
& $\Delta^\circ_{786}$ & $32.860.461$ & $20^6$ & $20^{12}$ & $51.3$ \\
& $\Delta^\circ_{762}$ & $32.858.151$ & $20^6$ & $20^{12}$ & $51.3$ \\
\cmidrule{2-6}
& $\Delta^\circ_{417}$ & $32.857.596$ & $20^6$ & $20^{12}$ & $51.3$ \\
& $\Delta^\circ_{838}$ & $32.845.047$ & $20^6$ & $20^{12}$ & $51.3$ \\
& $\Delta^\circ_{782}$ & $32.844.379$ & $20^6$ & $20^{12}$ & $51.3$ \\
& $\Delta^\circ_{377}$, $\Delta^\circ_{499}$, $\Delta^\circ_{503}$ & $30.846.440$ & $20^6$ & $20^{12}$ & $48.2$ \\
& $\Delta^\circ_{1348}$ & $30.845.702$ & $20^6$ & $20^{12}$ & $48.2$ \\
\cmidrule{2-6}
& $\Delta^\circ_{882}$, $\Delta^\circ_{856}$ & $30.840.098$ & $20^6$ & $20^{12}$ & $48.2$ \\
& $\Delta^\circ_{1340}$ & $28.954.543$ & $20^6$ & $20^{12}$ & $45.2$ \\
& $\Delta^\circ_{1879}$ & $28.950.852$ & $20^6$ & $20^{12}$ & $45.2$ \\
& $\Delta^\circ_{1384}$ & $27.178.020$ & $20^6$ & $20^{12}$ & $42.5$ \\
\bottomrule
\end{tabular}
\end{align}
The table presented here groups together polytopes whose dual graph of $C_{(\mathbf{3},\mathbf{2})_{1/6}}^\bullet$ are the same. Based on this initial outcome, it can be inferred that around 50\% of all limit root bundles have exactly three global sections, which is promising news for string phenomenology. It is worth reiterating that these counts are FRST-invariant, meaning they apply to the entire family of F-theory QSM geometries $B_3( \Delta^\circ )$.

The method for counting limit roots presented in \cite{BCL21} was expanded in \cite{BCDO22} to include the counting of global sections of specific partial blow-up limit roots. As an example, we can examine a limit root on a partial blow-up where we do not blow up the nodal singularity identified by the dashed edge:
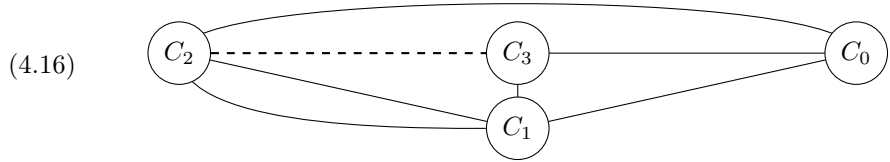
\begin{equation}
\begin{tikzpicture}[baseline=(current  bounding  box.center)]
      
      % define parameters
      \def\s{4.5};
      \def\h{1.0};
      
      % draw paths between curves
      \path[-, thick, dashed] (-\s,0) edge (0,0);
      \path[-] (-\s,0) edge (0,-\h);
      \path[-] (\s,0) edge (0,0);
      \path[-] (\s,0) edge (0,-\h);
      \path[-] (0,0) edge (0,-\h);
      \path[-, out = -90, in = 180, looseness = 0.6] (-\s,0) edge (0,-\h);
      \path[-, out = 90, in = 90, looseness = 0.25] (-\s,0) edge (\s,0);
      
      % draw curves
      \node at (1*\s,0) [circle, draw, fill=white]{$C_0$};
      \node at (0,-\h) [circle, draw, fill=white]{$C_1$};
      \node at (-1*\s,0) [circle, draw, fill=white]{$C_2$};
      \node at (0,0) [circle, draw, fill=white]{$C_3$};
      
\end{tikzpicture}
\label{equ:PartialBlowup}
\end{equation}
Such limit roots $P^\bullet$ correspond to line bundles on the following nodal curve:
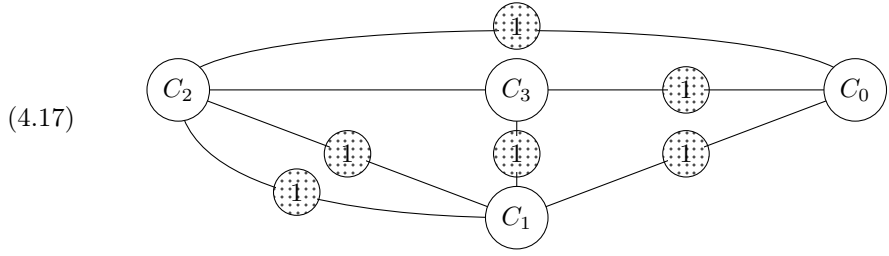
\begin{equation}
\begin{tikzpicture}[baseline=(current  bounding  box.center)]
      
      % define parameters
      \def\s{4.5};
      \def\h{1.7};
      
      % draw paths between curves
      \path[-] (-\s,0) edge (0,0);
      \path[-] (-\s,0) edge (0,-\h);
      \path[-] (\s,0) edge (0,0);
      \path[-] (\s,0) edge (0,-\h);
      \path[-] (0,0) edge (0,-\h);
      \path[-, out = -90, in = 180, looseness = 0.8] (-\s,0) edge (0,-\h);
      \path[-, out = 90, in = 90, looseness = 0.3] (-\s,0) edge (\s,0);
      
      % draw curves
      \node at (1*\s,0) [circle, draw, fill=white]{$C_0$};
      \node at (0,-\h) [circle, draw, fill=white]{$C_1$};
      \node at (-1*\s,0) [circle, draw, fill=white]{$C_2$};
      \node at (0,0) [circle, draw, fill=white]{$C_3$};
      
      % draw exceptional P1s
      \node at (0.5*\s,0*\h) [fill=white, scale = 2] {};
      \node at (0.5*\s,0*\h) [circle, draw, pattern=dots, pattern color = black!70]{$1$};
      
      \node at (0.5*\s,-0.5*\h) [fill=white, scale = 2] {};
      \node at (0.5*\s,-0.5*\h) [circle, draw, pattern=dots, pattern color = black!70]{$1$};
      
      \node at (-0.5*\s,-0.5*\h) [fill=white, scale = 2] {};
      \node at (-0.5*\s,-0.5*\h) [circle, draw, pattern=dots, pattern color = black!70]{$1$};
      
      \node at (0*\s,-0.5*\h) [fill=white, scale = 2] {};
      \node at (0*\s,-0.5*\h) [circle, draw, pattern=dots, pattern color = black!70]{$1$};
      
      \node at (0*\s,0.5*\h) [fill=white, scale = 2] {};
      \node at (0*\s,0.5*\h) [circle, draw, pattern=dots, pattern color = black!70]{$1$};
      
      \node at (-0.65*\s,-0.8*\h) [fill=white, scale = 2.3] {};
      \node at (-0.65*\s,-0.8*\h) [circle, draw, pattern=dots, pattern color = black!70]{$1$};
            
\end{tikzpicture}
\end{equation}
Once more, the sections on the exceptional $\mathbb{P}^1$s play a role in connecting the sections on neighboring components $C_i$. However, in this case, one node (marked by the dashed edge in \cref{equ:PartialBlowup}) has not been blown up, resulting in the number of global sections of $P^\bullet$ matching those of the restriction of $P^\bullet$ to the following nodal curve:
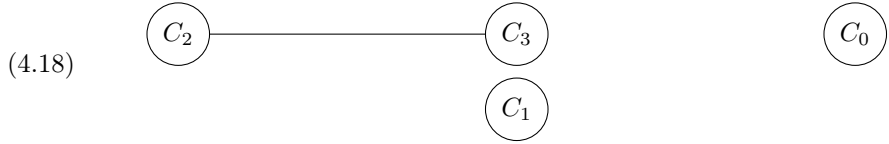
\begin{equation}
\begin{tikzpicture}[baseline=(current  bounding  box.center)]
      
      % define parameters
      \def\s{4.5};
      \def\h{1.0};
      
      % draw paths between curves
      \path[-] (-\s,0) edge (0,0);
      
      % draw curves
      \node at (1*\s,0) [circle, draw, fill=white]{$C_0$};
      \node at (0,-\h) [circle, draw, fill=white]{$C_1$};
      \node at (-1*\s,0) [circle, draw, fill=white]{$C_2$};
      \node at (0,0) [circle, draw, fill=white]{$C_3$};
      
\end{tikzpicture}
\end{equation}
It should be noted that $C_2 \cup C_3$ forms a tree-like nodal curve. In \cite{BCDO22}, an algorithm was developed for computing the cohomology of line bundles on such curves, which allowed us to determine the number of global sections using the formula:
\begin{align}
h^0( C^\bullet, P^\bullet) = h^0 \left( C_0, \left. P^\bullet \right|_{C_0} \right) + h^0 \left( C_1, \left. P^\bullet \right|_{C_1} \right) + h^0 \left( C_2 \cup C_3, \left. P^\bullet \right|_{C_2 \cup C_3} \right) \, .
\end{align}
Our computer implementation determined the number of limit roots for $\Delta_4^\circ$ based on the remaining number of nodes, denoted as $N$:
\begin{align}
\begin{tabular}{c|cc|cc}
\toprule
 & \multicolumn{2}{c|}{Percentages} & \multicolumn{2}{c}{Absolute numbers} \\
$N$ & $h^0 = 3$ & $h^0 \geq 3$ & $h^0 = 4$ & $h^0 \geq 4$ \\
\midrule
0 & 53.6 & & 11110 & \\
1 & 36.7 & & 7601 & \\
2 & 7.5 & 0.5 & 1562 & 110 \\
3 & 1.3 & 0.1 & 264 & 11 \\
4 & & 0.3 & & 66 \\
5 & & 0.1 & & 11 \\
7 & & 0.0 & & 1 \\
\midrule
$\Sigma$ & 99.0 & 1.0 & 20537 & 199 \\
\bottomrule
\end{tabular} \label{equ:DetailedCount}
\end{align}
The absolute numbers are expressed as multiples of $\mu = 12^4$. For example, there are $7601 \times 12^4$ 12-th limits roots of $\overline{K}_{C_{(\mathbf{3},\mathbf{2})_{1/6}}^\bullet}^{12}$ with $h^0 = 3$, resulting from not blowing up precisely $N = 1$ nodes. The algorithm covered all limit roots, as evidenced by $12^4 \cdot \left( 20537 + 199 \right) = 12^8$. Almost 99\% of all limit roots have precisely three sections on $C_{(\mathbf{3},\mathbf{2})_{1/6}}^\bullet$, which is a significant improvement over the findings of \cite{BCL21}. This procedure can be repeated for each of the selected 33 QSM polytopes \cite{BCDO22}.
\begin{align}
\begin{tabular}{c|cc|cc|cc|cc}
\toprule
Polytope & $= 3$ & $\geq 3$ & $= 4$ & $\geq 4$ & $= 5$ & $\geq 5$ & $= 6$ & $\geq 6$ \\
\midrule
$\Delta_8^\circ$ & 76.4 & 23.6 & & & & \\
$\Delta_4^\circ$ & 99.0 & 1.0 & & & & \\
$\Delta_{134}^\circ$ & 99.8 & 0.2 & & & & \\
$\Delta_{128}^\circ$, $\Delta_{130}^\circ$, $\Delta_{136}^\circ$, $\Delta_{236}^\circ$ & 99.9 & 0.1 & & & & \\
\midrule \midrule
$\Delta_{88}^\circ$ & 74.9 & 22.1 & 2.5 & 0.5 & 0.0 & 0.0 \\
$\Delta_{110}^\circ$&82.4&14.1&3.1&0.4&0.0 & \\
$\Delta_{272}^\circ$, $\Delta_{274}^\circ$&78.1&18.0&3.4&0.5&0.0&0.0 \\
$\Delta_{387}^\circ$ & 73.8 & 21.9 & 3.5 & 0.8 & 0.0 & 0.0 \\
$\Delta_{798}^\circ$, $\Delta_{808}^\circ$, $\Delta_{810}^\circ$, $\Delta_{812}^\circ$ & 77.0 & 17.9 & 4.4 & 0.7 & 0.0 & 0.0 \\
\midrule
$\Delta_{254}^\circ$ &95.9&0.5&3.5&0.0&0.0&0.0 \\
$\Delta_{52}^\circ$ &95.3&0.7&3.9&0.0&0.0&0.0 \\
$\Delta_{302}^\circ$ &95.9&0.5&3.5&0.0&0.0 &\\
$\Delta_{786}^\circ$ &94.8&0.3&4.8&0.0&0.0&0.0 \\
$\Delta_{762}^\circ$ &94.8&0.3&4.9&0.0&0.0&0.0 \\
\midrule
$\Delta_{417}^\circ$ & 94.8&0.3&4.8&0.0&0.0&0.0&0.0 \\
$\Delta_{838}^\circ$ &94.7&0.3&5.0&0.0&0.0&0.0 \\
$\Delta_{782}^\circ$ &94.6&0.3&5.0&0.0&0.0&0.0 \\
$\Delta_{377}^\circ$, $\Delta_{499}^\circ$, $\Delta_{503}^\circ$ & 93.4 & 0.2 & 6.2 & 0.0 & 0.1 & 0.0 \\
$\Delta_{1348}^\circ$ &93.7&0.0&6.2&0.0&0.1&&0.0 \\
\midrule
$\Delta_{882}^\circ$, $\Delta_{856}^\circ$&93.4&0.3&6.2&0.0&0.1&0.0&0.0 \\
$\Delta_{1340}^\circ$ & 92.3&0.0&7.6&0.0&0.1&&0.0 \\
$\Delta_{1879}^\circ$ & 92.3&0.0&7.5&0.0&0.1&&0.0 \\
$\Delta_{1384}^\circ$ &90.9&0.0&8.9&0.0&0.2&&0.0 \\
\bottomrule
\end{tabular} \label{tab:Results1}
\end{align}
Based on these findings, we can infer that the majority of (partial) limit root bundles have exactly three global sections. This is excellent news for string phenomenology, as it implies that it is highly probable that there are no vector-like exotics on the quark-doublet matter curve.

The observant reader may have noticed that our results are considerably more conclusive for some polytopes than for others. For example, we establish a lower bound for $h^0$ in roughly 23\% of the limit roots for $\Delta_8^\circ$, while for $\Delta_4^\circ$, we cannot determine $h^0$ for only about 1\% of the limit roots. The primary factor influencing this discrepancy is the presence of an elliptic curve component in $\Delta_8^\circ$. When the nodal quark-doublet curve contains at least one elliptic curve, we face limitations with partial blow-up root bundles. Firstly, the line bundle cohomology technology on tree-like nodal curves presented in \cite{BCDO22} is no longer applicable if the elliptic curve has one unblown node. Secondly, if there is a $d=0$ line bundle on an elliptic curve, and we are searching for $r$-th roots, then $r^2-1$ of these roots are guaranteed to lack global sections. The number of global sections in the one remaining root is determined by whether the original bundle was trivial or not, and distinguishing between the two situations is extremely challenging. Therefore, we can only provide a lower bound in such cases. Indeed, all entries in \cref{tab:Results1} that indicate an uncertainty greater than 1\% contain an elliptic curve.

It is intriguing to compare these results to the classical Brill-Noether theory, which deals with a continuous space of line bundles and estimates the dimension of the variety of line bundles with a specific $h^0$. However, this notion does not apply to root bundles since there are only finitely many of them. To the best of our knowledge, \cref{tab:Results1} (further information in \cite{BCDO22}) represents the first counts/estimates for such bundles. As a result, we propose that these tables be regarded as an initial response to the question ``What is Brill-Noether theory for limit roots?''. In particular, we notice that for the QSM configurations with $\overline{K}_{B_3}^3 = 10$, we find limit root bundles that always have more than the minimum number of global sections.

\subsection{Probability for absence of vector-like exotics in $\mathbf{B_3( \Delta_4^\circ )}$} \label{subsec:D4}

To gain a complete understanding, it is important to investigate the limit roots for which the methods mentioned earlier were unable to determine the number of global sections. Therefore, we will now examine the unresolved cases for the polytope $\Delta_4^\circ$. There are only five configurations to be discussed after taking symmetry into account. We will begin with the first configuration:
\begin{equation}
\begin{tikzpicture}[scale=0.6, baseline=(current  bounding  box.center)]
      
      % define parameters
      \def\s{3.5};
      
      % draw paths between curves
      \path[-,out = 45, in = 135, looseness = 0.6] (0,0) edge (2*\s,0);
      \path[-,out = -45, in = -135, looseness = 0.6] (0,0) edge (2*\s,0);
      
      % draw curves
      \node at (-2*\s,0) [circle, draw, fill=white, label=left:$C_1$]{$d = 0$};
      \node at (0,0) [circle, draw, fill=white, label=left:$C_2$]{$d = 1$};
      \node at (2*\s,0) [circle, draw, fill=white, label=right:$C_3$]{$d = 1$};
      
\end{tikzpicture} \label{equ:StationaryCircuit1}
\end{equation}
A line bundle on a nodal curve whose irreducible components are all isomorphic to $\mathbb{P}^1$ can be uniquely determined by two pieces of information as stated in \cite{HM06}: the degree of the line bundle on each irreducible component, and the descent data which specifies the gluing conditions at the nodes. The descent data is represented by a number $\lambda \in \mathbb{C}^\ast$ for each node, and the global sections at each node must be equal up to this factor $\lambda$. One approach to investigate the unresolved cases for the polytope $\Delta_4^\circ$ is to trace the descent data through the limit root construction of \cite{CCC07}. Alternatively, we can study all different descent data that can be put on the nodal curve represented by \cref{equ:StationaryCircuit1}, i.e., all line bundles on the curve with the given multidegree. We choose the second approach, as every such line bundle has exactly two global sections on $C_2 \cup C_3$. The tracing of descent data through the limit root prescription of \cite{CCC07} will be left for future work. We select coordinates for $C_2$ and $C_3$ and parametrize the local sections on $C_2$, $C_3$ using $\left( \alpha_1, \alpha_2, \alpha_3, \alpha_4 \right) \in \mathbb{C}^4$:
\begin{align}
\begin{tabular}{ccc}
\toprule
Curve & Coordinates & Sections \\
\midrule
$C_2$ & $[a:b]$ & $\alpha_1 a + \alpha_2 b$ \\
$C_3$ & $[c:d]$ & $\alpha_3 c + \alpha_4 d$ \\
\bottomrule
\end{tabular}
\end{align}
By use of a M\"obius transformation, we place the nodes at the following positions:
\begin{align}
\begin{tabular}{c|cc}
\toprule
Label & Coordinates in $C_2$ & Coordinates in $C_3$ \\
\midrule
$n_1$ & $[a:b] = [1:0]$ & $[c:d] = [1:0]$\\
$n_2$ & $[a:b] = [0:1]$ & $[c:d] = [0:1]$\\
\bottomrule
\end{tabular}
\end{align}
Next, let us enforce the gluing conditions:
\begin{align}
n_1 \colon \qquad & \left( \alpha_1 a + \alpha_2 b \right)([1:0]) = \lambda_1 \cdot \left(\alpha_3 c + \alpha_4 d\right)([1:0]) \, , \qquad \lambda_1 \in \mathbb{C}^\ast \, , \\
n_2 \colon \qquad & \left( \alpha_1 a + \alpha_2 b \right)([0:1]) = \lambda_2 \cdot \left(\alpha_3 c + \alpha_4 d\right)([0:1]) \, , \qquad \lambda_2 \in \mathbb{C}^\ast \, .
\end{align}
Hence, the global sections on $C_2 \cup C_3$ are parametrized by $(\alpha_3, \alpha_4) \in \mathbb{C}^2$ via
\begin{align}
\begin{tabular}{ccc}
\toprule
Curve & Coordinates & Sections \\
\midrule
$C_2$ & $[a:b]$ & $\lambda_1 \alpha_3 a + \lambda_2 \alpha_4 b$ \\
$C_3$ & $[c:d]$ & $\alpha_3 c + \alpha_4 d$ \\
\bottomrule
\end{tabular}
\end{align}
By also considering $C_1$, we find that the number of global sections for the configuration in \cref{equ:StationaryCircuit1} is three. It is worth noting that we could have rescaled the sections on $C_2$ to eliminate one of the $\lambda$ parameters, resulting in only one remaining parameter, which is equivalent to the first Betti number of the graph \cite{HM06}. We can perform a similar analysis for three other graphs of this type, all of which also have $h^0( C^\bullet, P^\bullet ) = 3$ \cite{BCDO22}. This improves the original Brill-Noether table in \cref{equ:DetailedCount}:
\begin{align}
\begin{tabular}{c|cc|cc}
\toprule
& \multicolumn{2}{c|}{Percentages} & \multicolumn{2}{c}{Absolute numbers} \\
$N$ & $h^0 = 3$ & $h^0 \geq 3$ & $h^0 = 3$ & $h^0 \geq 3$ \\
\midrule
0 & 53.6 &  & 11110 &  \\
1 & 36.7 &  & 7601 &  \\
2 & 8.0 &  & 1672 &  \\
3 & 1.3 &  & 275 &  \\
4 & 0.3 &  & 66 &  \\
5 & 0.1 &  & 11 &  \\
7 &  & 0.0 &  & 1 \\
\midrule
$\Sigma$ & 100.0 & 0.0 & 20735 & 1 \\
\bottomrule
\end{tabular}
\end{align}
Recall from \cref{tab:RootBundles} that we are looking for limit 12th roots of the 12th power of $K_{C_{(\mathbf{3}, \mathbf{2})_{1/6}}^\bullet}$. It is evident that $K_{C_{(\mathbf{3}, \mathbf{2})_{1/6}}^\bullet}$ is a solution to this root bundle equation and that the number of global sections of the canonical bundle always matches the arithmetic genus, which is four in this case. Since all the limit roots found above had $h^0 = 3$, this final configuration must be $K_{C_{(\mathbf{3}, \mathbf{2})_{1/6}}^\bullet}$ and have $h^0 = 4$. Therefore, only more than $99.995\%$ of the limit root bundles on the nodal curves $C^\bullet_{(\mathbf{\overline{3}}, \mathbf{2})_{1/6}}$ have exactly three global sections. It is possible to lose sections along $C^\bullet_{(\mathbf{\overline{3}}, \mathbf{2})_{1/6}} \to C_{(\mathbf{\overline{3}}, \mathbf{2})_{1/6}}$, which means that -- statistically speaking -- the probability of exotic vector-like quark-doublets being absent in the family $B_3( \Delta_4^\circ )$ is higher than $99.995\%$.

\section{Summary and Outlook} \label{sec:SummaryAndOutlook}

Studies on vector-like spectra of 4d F-theory compactifications often feature root bundles. This is particularly true for the Quadrillion F-theory Standard Models \cite{CHLLT19} for which zero modes in the representations $(\mathbf{3}, \mathbf{2})_{1/6}$, $(\mathbf{\overline{3}}, \mathbf{1})_{-2/3}$, and $(\mathbf{1}, \mathbf{1})_{1}$ are counted by root bundles. A superset of these root bundles has been studied extensively in \cite{BCDLO21, BCL21, BCDO22}. In particular, \cite{BCDO22} focused on the family $B_3( \Delta_4^\circ )$ of $\mathcal{O}(10^{11})$ F-theory QSM. The geometries of these F-theory vacua are defined by toric 3-folds which are defined by full, regular, star triangulations of the 4-th polytope in the Kreuzer-Skarke list of 3-dimensional, reflexive poytopes \cite{KS98}. It was found that more than $99.995\%$ of the root bundles in the superset of studied root bundles have precisely three global sections. This discovery resulted from a thorough investigation of a significant portion of QSM geometries. Although the outcomes are most conclusive for the family $B_3( \Delta_4^\circ )$, the findings for this extensive subset of QSM geometries offer statistical evidence supporting the possibility of the non-existence of vector-like exotics in the representations $(\mathbf{3}, \mathbf{2})_{1/6}$, $(\mathbf{\overline{3}}, \mathbf{1})_{-2/3}$ and $(\mathbf{1}, \mathbf{1})_{1}$ within the F-theory QSMs.

In QSM geometries, vector-like spectra are calculated by counting the cohomologies of line bundles, which are roots of the (twists of certain powers of) canonical bundle \cite{BCDLO21}. These root bundles $P_{\mathbf{R}}$ on the matter curve $C_{\mathbf{R}}$ are not unique, and it is unclear which ones are induced from F-theory gauge potentials in the Deligne cohomology. Investigating this matter is a complex task beyond our current abilities. Therefore, the studies in \cite{BCDLO21, BCL21, BCDO22} analyzed \emph{all} mathematically permissible root bundles, including those that could potentially be induced from the $G_4$-flux and all spin bundles on the matter curve under examination. This study is limited to a local and bottom-up analysis of one matter curve at a time, neglecting any correlations among the vector-like spectra on different matter curves. Consequently, a (proper) superset of all physically acceptable root bundles was examined.

To count the root bundles on the matter curve $C_{\mathbf{R}}$, it is helpful to deform the curve into a nodal curve. This deformation provides two advantages. Firstly, constructing root bundles on smooth curves is challenging, while constructing them on nodal curves using limit roots has been extensively described and involves combinatorial tasks \cite{CCC07}. Secondly, for each family $B_3( \Delta^\circ )$ of F-theory QSMs, there is a canonical nodal curve $C_{\mathbf{R}}^\bullet$ that solely depends on $\Delta^\circ$. Thus, by examining limit roots on this canonical nodal curve $C_{\mathbf{R}}^\bullet$, one can estimate the vector-like spectra for the entire family $B_3( \Delta^\circ )$ with a single calculation.

This method necessitates the capability to enumerate limit root bundles with specific numbers of global sections, which is precisely within the realm of Brill-Noether theory. For F-theory QSMs, the global sections of most limit roots can be calculated by finding the cohomologies of line bundles on tree-like rational curves, which can be achieved with a computer algorithm described in \cite{BCDO22}. However, for some limit roots, line bundle cohomology on circuit-like curves is needed, which is currently being researched. This is not a big issue for F-theory QSMs because the nodal curves being studied have relatively small Betti numbers and high root indices, resulting in few contributions from circuits. If the root indices were lower, circuits would become more significant, and for second roots, circuit-like and tree-like graphs would contribute equally, with circuits potentially dominating.

The algorithm used in \cite{BCDO22} simplifies a pair consisting of a nodal curve $C^\bullet$ and a line bundle $L$ on it by removing components of $C^\bullet$ and adjusting $L$ in such a way that the number of global sections remains the same. This process ends with a final configuration, from which the number of global sections can be easily determined. For partial blow-up limit roots that go beyond the tree-like case, the counting of global sections had to be performed manually. In the case of the $B_3(\Delta_4^\circ)$ family, this involved studying line bundle cohomologies on no more than five nodal, circuit-like curves. This study revealed that on four of these nodal curves, the line bundle had exactly three sections. The remaining case, denoted as a ``jumping circuit'' in \cite{BCDO22}, was identified as the canonical bundle, which must have as many sections as the arithmetic genus of the nodal case, which turned out to be four. This is why ``only'' more than $99.995\%$ of limit root bundles on the nodal curves $C_{(\mathbf{3}, \mathbf{2})_{1/6}}^\bullet$, $C_{(\mathbf{\overline{3}}, \mathbf{1})_{-2/3}}^\bullet$, and $C_{(\mathbf{1}, \mathbf{1})_{1}}^\bullet$ have exactly three global sections.

The ignorances in \cref{tab:Results1} can be attributed to two factors. Firstly, when a node remains on an elliptic curve, the methods used in \cite{BCDLO21, BCL21, BCDO22} only yield a lower bound. Secondly, at the time of the publication of \cite{BCDO22}, there was no algorithmic method for computing the line bundle cohomology on circuit-like nodal curves, which meant that even in these cases, only a lower bound was obtained. The employed computer algorithm can be found in \cite{Bie23}. Currently, this algorithm is being extended to include functionality to compute line bundle cohomology on circuit-like nodal curves and to provide better lower bounds. Interested readers may be tempted to perform the relevant computations themselves using a personal computer. The results at the top of \cref{tab:Results1} should take only a few minutes to compute while those at the bottom may take more than one day.

The table in \cref{tab:Results1} demonstrates that certain limit roots on nodal matter curves with $\overline{K}_{B_3}^3 = 10$ possess more than three global sections. Consequently, it is necessary to establish a set of conditions that maintain the number of sections or reduce it as the matter curve is smoothed out, especially to ensure the presence of a single Higgs pair. One possible avenue to identifying these conditions is to examine Yukawa interactions, which provide information about which fields acquire mass, when they acquire it, and how they do so during deformation. Insights into these conditions could be gained by conducting a similar investigation to that described in \cite{CLLZZ20}. Future research on explicit realizations of F-theory MSSMs will likely need to address this issue.

The analysis presented in this paper focuses on the vector-like spectra of F-theory QSMs for particle phenomenology. However, it also represents the -- to the best knowledge of the author -- \emph{first} computational approach to Brill-Noether theory of limit root bundles. While classical Brill-Noether theory is concerned with the classification of a continuous space of line bundles by their global sections, the number of root bundles in this case is finite and denoted as $N_{\text{total}}$. Thus, the Brill-Noether theory in this context involves finding a partition $N_{\text{total}} = N_0 + N_1 + \dots$, where $N_i$ represents the number of root bundles with $i$ global sections. The main results presented in \cref{tab:Results1} can be interpreted in this way.

The approach presented in this paper for studying the Brill-Noether theory of limit root bundles may have potential applications in machine learning. By using the algorithm \cite{Bie23}, an approximation of the Brill-Noether theory of limit roots for a collection of simple graphs can be computed (modulo the mentioned caveats). It would be interesting to investigate if there is a pattern linking the graphs to the corresponding Brill-Noether approximations. This could provide mathematical insights. Moreover, a well-trained algorithm could be used for phenomenological applications, such as estimating the limit root bundles on the Higgs curve in F-theory QSMs, which are currently beyond our computational capabilities.

To conclude, it should be noted that the previous studies \cite{BCDLO21, BCL21, BCDO22} did not address the fundamental question of which mathematical roots are induced by F-theory gauge potentials. It is likely that answering this question is complex, and so has not yet been attempted. Nevertheless, even though more than 99.995\% of limit root bundles in representation $(\mathbf{3}, \mathbf{2})_{1/6}$ in the family of F-theory QSMs $B_3( \Delta_4^\circ )$ have exactly three global sections, the definitive answer regarding the presence or absence of vector-like exotics can only be obtained by determining which root bundles can be realized by F-theory gauge potentials. The detailed examination of such top-down conditions is planned for future research.

\subsection*{Acknowledgement}
M.~B. expresses gratitude towards his colleagues Mirjam Cveti\v{c}, Ron Donagi, Muyang Liu and Marielle Ong for their contributions and valuable discussions. This work was supported by the SFB-TRR 195 \emph{Symbolic Tools in Mathematics and their Application} of the German Research Foundation (DFG). The author is also grateful for finanical support by the Forschunginitiative des Landes Rheinland-Pfalz through \emph{SymbTools – Symbolic Tools in Mathematics and their Application}.

\bibliographystyle{amsplain}
\bibliography{references}{}

\end{document}